\documentclass[%
 reprint,
 amsmath,amssymb,
 aps,
]{revtex4-2}

\usepackage{hyperref}       
\hypersetup{
    colorlinks=true,
    linkcolor=blue,
    filecolor=magenta,      
    urlcolor=cyan,
}

\usepackage{graphicx}
\usepackage{dcolumn}
\usepackage{bm}

\newcommand{\abs}[1]{\left|#1\right|}
\newcommand{\abssq}[1]{\abs{#1}^2}
\newcommand{\AS}{A_\mathrm{s}}
\newcommand{\AAS}{A_\mathrm{as}}
\newcommand{\ASmu}{A_{\mathrm{s},\mu}}

\newcommand{\AASmu}{A_{\mathrm{as},\mu}}

\DeclareMathOperator{\FFT}{\mathcal{F}}
\renewcommand{\Re}{\text{Re}}
\renewcommand{\Im}{\text{Im}}

\newcommand{\Jc}{J} 
\newcommand{\gc}{g_K} 
\newcommand{\kap}[1]{\kappa_{\mathrm{ex},#1}}

\newcommand{\sym}{\mathrm{s}}
\newcommand{\opa}{\hat{a}}

\newcommand{\ant}{\mathrm{as}}

\graphicspath{{./Figs/}}

\begin{document}

\preprint{APS/123-QED}

\title{Dissipative Kerr solitons in a photonic dimer\texorpdfstring{\\}{} on both sides of exceptional point}

\author{K.~Komagata$^{1,2}$}
\author{A.~Tikan$^{1}$}
\email{alexey.tikan@epfl.ch}
\author{A.~Tusnin$^{1}$}
\author{J.~Riemensberger$^{1}$}
\author{M.~Churaev$^{1}$}
\author{H.~Guo$^{1,3}$}
\author{T.J.~Kippenberg$^1$}
\email{tobias.kippenberg@epfl.ch}

\affiliation{$^1$Institute of Physics, Swiss Federal Institute of Technology Lausanne (EPFL), CH-1015 Lausanne, Switzerland \\ $^2$Present address: Laboratoire Temps-Fréquence, Institut de Physique, Université de Neuch\^atel, CH-2000 Neuch\^atel, Switzerland \\
$^3$Present address: Key Laboratory of Specialty Fiber Optics and Optical Access Networks, Shanghai University, 200444 Shanghai, China}

\date{\today}

\begin{abstract}
Exceptional points are a ubiquitous concept widely present in driven-dissipative coupled systems described by a non-Hermitian Hamiltonian. It is characterized by the degeneracy of the Hamiltonian's eigenvalues and coalescence of corresponding eigenvectors. Recent developments demonstrated that exceptional points can play an important role in photonics. However, to date, exceptional points have been extensively examined in the systems supporting only a few optical modes, thereby leaving the observation of collective (multimode) effects outside of the scope of study. In the present paper, we analyze the role of exceptional points in nonlinear multimode photonics. Specifically, we provide insights into complex nonlinear dynamics arising in a continuous wave-driven pair of strongly coupled nonlinear micro-resonators (i.e. a nonlinear photonic dimer) operating in the multimode regime. Investigating this system, which is known to possess exceptional points, we find two fundamentally different nonlinear regimes of operation corresponding to effective parity-time symmetric and broken parity-time symmetry states. We demonstrate that the photonic dimer can be critically coupled to a bus waveguide, thereby, providing an efficient generation of the dissipative Kerr solitons on both sides of the exceptional point. The parity-time symmetric case, which corresponds to a pair of symmetrically split resonances, has been recently shown to exhibit a variety of emergent phenomena including gear soliton generation, symmetry breaking, and soliton hopping. Dissipative solitons generation in the parity-time symmetry broken case - leading to the dissipation splitting - up to now remains unexplored.
\end{abstract}

\maketitle

\section{\label{sec:0}  Introduction}

Symmetry is one of the main fundamental concepts in physics which underpins conservation laws, micro- and macro-properties of matter, rising of degeneracies, etc. Breaking of symmetry in spontaneous or deterministic way is responsible for a variety of phenomena with a particularly illustrative example of the Higgs mechanism explaining the appearance of masses of certain bosons~\cite{higgs2014nobel}. Systems obeying Parity-Time ($\mathcal{PT}$) symmetry have been shown to provide a possibility to treat open quantum systems described by a non-Hermitian Hamiltonian and nonetheless retrieve a real spectrum of eigenvalues~\cite{Bender1998Real,Ganainy2018Hermitian}. Eigenvalues of $\mathcal{PT}$ symmetric Hamiltonian have two typical regions on the parameters space corresponding to preserved and broken symmetry. The transient point where the eigenvalues and eigenvectors coalesce is called an exceptional point (EP)~\cite{Miri2019Exceptional,Ozdemir2019Parity}.

The effect has been observed in various physical systems~\cite{Ganainy2018Hermitian}. Due to the well-controlled laboratory conditions and a wide range of possible applications, guided optics serves as one of the primary platforms for investigating effects that emerge in $\mathcal{PT}$-symmetric systems~\cite{Ruter2010Observation,Longhi2017,Feng2017Hermitian,Ozdemir2019Parity}. These effects cover observation of strong non-reciprocity in banded waveguides, enhanced lasing and (classical noise limited) sensing in coupled ring resonators with loss and gain (for more examples see review article~\cite{Ozdemir2019Parity}). Remarkably, after a gauge transformation, $\mathcal{PT}$ symmetry can be unraveled in completely passive resonators having different loss rates~\cite{peng2014loss,Ozdemir2019Parity}. Here, it is important to mention that the majority of these studies have been considering linear effects in single-mode arrangements or nonlinear ones in non-resonant systems~\cite{Konotop2016Nonlinear}.

Almost in parallel, and without apparent intersection, another field of study has been rapidly developing in photonics: design and fabrication of passive nonlinear coherent broadband light sources~\cite{Gaeta2019Photonic}. For this purpose, exactly the same optical platforms have been used: optical waveguides and resonators but in the opposite, strongly nonlinear regime. Nonlinear waveguides served as a sources of supercontinuum signals~\cite{guo2018mid}, while micro- and macro-resonators has been used for generation of stable and coherent frequency combs~\cite{Kippenberg2018Dissipative}. The latter has been achieved due to the observation of localized coherent structures in passive optical micro-resonators~\cite{Herr2014Temporal}. These structures generated in media with $\chi^{(3)}$ nonlinearity are called dissipative Kerr solitons (DKSs). They have been widely investigated in nonlinear photonics over the last decades~\cite{wabnitz1993suppression,barashenkov1996existence,leo2010temporal,Kippenberg2018Dissipative}. The stability of DKSs relies on the balance between chromatic dispersion, Kerr-type nonlinearity, parametric gain and the intrinsic cavity losses~\cite{Herr2014Mode}. DKSs are exact solutions of the damped-driven nonlinear Schr\"odinger equation known as Lugiato-Lefever equation (LLE)~\cite{Haelterman1992Dissipative,Chembo2013Spatiotemporal}. Their observation in passive macroscopic (fiber) cavities~\cite{leo2010temporal} and (integrated) microcavity systems~\cite{Herr2014Temporal,brasch2016photonic} has spurred a vivid research effort unravelling a rich inherent dynamical behaviour~\cite{jang2015temporal,guo2017universal,Lucas2017Breathing,wang2017universal,Guo2017Intermode,karpov2019dynamics,yu2020spontaneous}. The discovery of DKSs in microresonators revolutionized the field, bringing coherent frequency combs to outside-of-laboratory applications~\cite{Kippenberg2018Dissipative}.
Remarkably, a pioneering theoretical proposal considering the DKS generation in the $\mathcal{PT}$-symmetric system has been presented in~\cite{milian2018cavity}.

Recently, DKSs have been discovered in a high-Q multimode photonic dimer (pair of strongly-coupled, almost identical nonlinear resonators)~\cite{tikan2020emergent}. The photonic dimer has revealed a pleiad of emergent phenomena including soliton hopping, periodic appearance of commensurate and incommensurate dispersive waves (DWs), and symmetry breaking related to the discreteness of the system. Solitons have been generated in both resonators simultaneously and due to the underlying field symmetry were called \textit{gear solitons} (GSs).

In this work, we investigate all-passive photonic dimer with a hidden $\mathcal{PT}$-symmetry (\textit{further referred to as $\mathcal{PT}$-symmetry for simplicity}) in linear and nonlinear multimode regimes. In the linear regime, we analyse conditions for the critical coupling and demonstrate that the line of EPs is a demarcation of these conditions. The exceptional point line splits the parameter space into two parts which we refer to as split resonance ($\mathcal{PT}$ symmetric) and split dissipation ($\mathcal{PT}$-symmetry broken). In the split resonance regime, we show novel insights into the effects previously reported in~\cite{tikan2020emergent}, using the supermode basis representation. Passing through an EP, we observe the divergence of the nonlinear interaction efficiency which hints at the enhanced sensitivity. Further, we investigate the nonlinear dynamics in the split dissipation regime which includes single-resonator DKS, dark-bright DKSs pairs, and highly-efficient perfect soliton crystals. 
Finally, we demonstrate switching of the soliton-generating cavity caused by the nonlinear alteration of the $\mathcal{PT}$ symmetry.
\section{\label{sec:1} Exceptional point as a demarcation of the critical coupling conditions}
\begin{figure*}
    \centering
    \includegraphics[width=\linewidth]{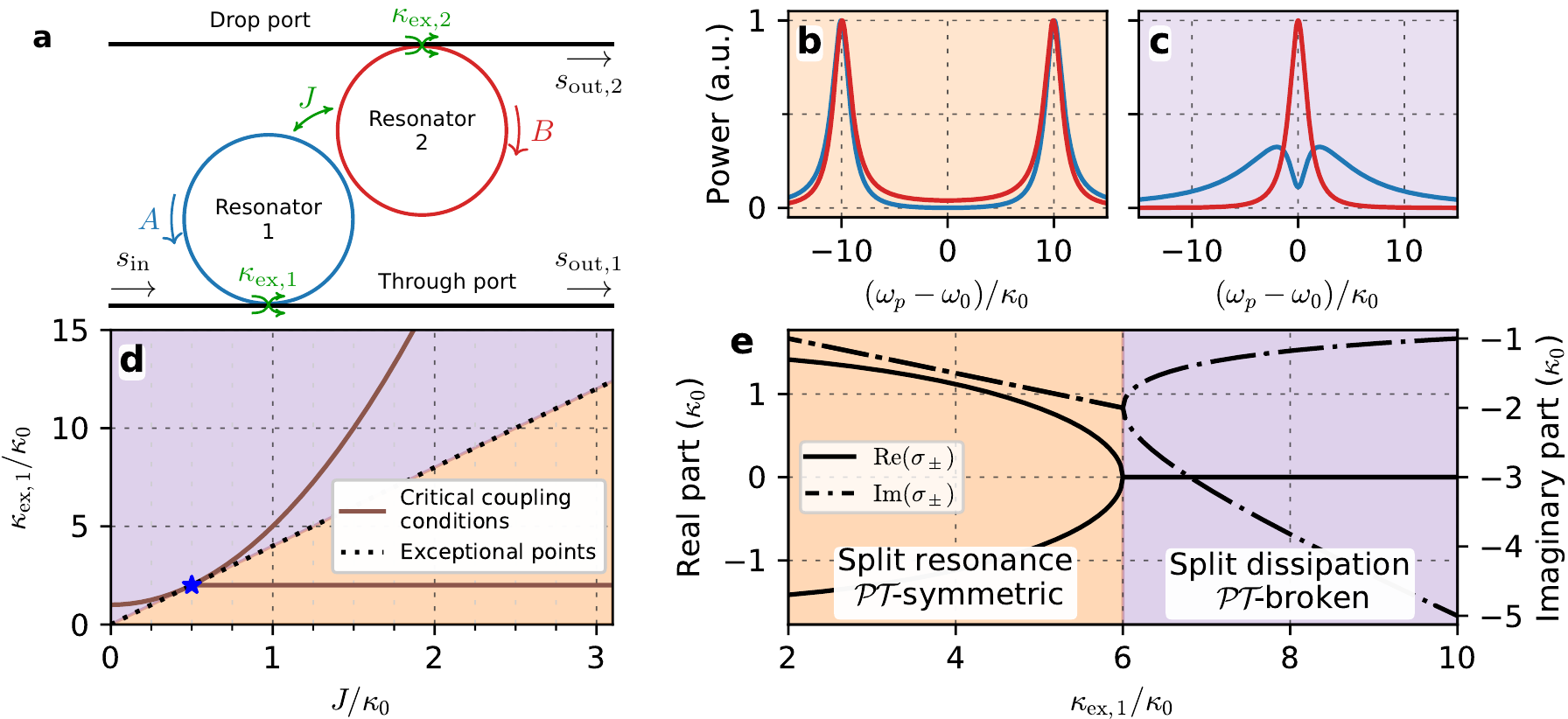}
    \caption{\textbf{Linear analysis of the photonic dimer.} (a) Schematic of the two resonators. (b,c) Field intensity in the resonator 1 (blue) and 2 (red) as a function of the laser detuning. (b) Shows a typical resonance splitting at critical coupling for $J = 10\kappa_0$, (c) the same for the split dissipation regime at critical coupling for $J = 1.5\kappa_0 $. (d) Critical coupling coefficients as a function of $J$ for $\kap2 = 0$, in the split dissipation (resonances) regime above (below) the exceptional point line (dotted). The two conditions branch off at the exceptional point of $J=1/2\kappa_0$, highlighted by the star. (e) Real and imaginary parts of the eigenvalues as a function of $\kap1$ for $J = 1.5\kappa_0$.}
    \label{fig:criticalCouplingSimple}
\end{figure*}

We consider the system of two multimode resonators [Fig.~\ref{fig:criticalCouplingSimple}(a)], with identical intrinsic loss rate $\kappa_0$, mode spacing $D_1$, and geometry, such that the dispersion and Kerr nonlinearity coefficient are also identical. A global offset between the resonant frequencies $\omega_\mu$ of their respective modes $\mu$ is introduced with the inter-resonator detuning $\delta$. The two resonators are coupled to each other by the evanescent field with the rate $\Jc_\mu$, which generally depends on the mode number. Each resonator is coupled to a waveguide (through and drop ports) with the rates $\kap{i},~i=1,2$. Resonator 1 is pumped by a continuous wave (cw) laser at frequency $\omega_p$. Nonlinear dynamics in the photonic dimer can be described by two coupled Lugiato-Lefever equations (LLEs), which in Fourier space is expressed as follows \cite{Hansson2014numerical,tikan2020emergent}:
\begin{align}
\frac{\mathrm{d}}{\mathrm{d}t} A_\mu &= -[\tfrac{1}{2}(\kappa_0 + \kappa_\mathrm{ex,1}) + i (\omega_\mu +\tfrac{1}{2}\delta - \mu D_1- \omega_p)]A_\mu  \notag\\
&\quad+ i \gc \FFT[A \abssq{A}]_\mu  + i \Jc_\mu B_\mu\notag+ \delta_{\mu,0}\sqrt{\kappa_\mathrm{ex,1}}s_\mathrm{in} \\
\frac{\mathrm{d}}{\mathrm{d}t} B_\mu &= -[\tfrac{1}{2}(\kappa_0 + \kappa_\mathrm{ex,2}) + i (\omega_\mu -\tfrac{1}{2}\delta- \mu D_1- \omega_p)]B_\mu  \notag\\
&\quad+ i \gc \FFT[B \abssq{B}]_\mu  + i \Jc_\mu A_\mu,
\label{eq:cLLEs}
\end{align}
where $g_\mathrm{K} = \frac{\hbar \omega_0^2 c n_2}{n_0^2 V_{\mathrm{eff}}}$ is the Kerr coefficient, $c$ stands for the speed of light in vacuum,  $\hbar$ - the Planck constant, $\omega_0$ - the frequency of the pumped mode, $V_{\mathrm{eff}}$ - the effective mode volume, $n_0$ and $n_2$ are linear and nonlinear refractive indexes, respectively, $\delta_{\mu,0}$ is the Kronecker delta, $s_\mathrm{in}=\sqrt{\frac{P_\mathrm{in}}{\hbar\omega_0}}$ - the input pump field amplitude, $A_\mu,~B_\mu$ are the field amplitudes of the modes with index $\mu$ in the first and second resonator, respectively. The variables $A,~B$ are the slowly varying intra-resonator field envelops, and $\FFT[...]_\mu$ denotes the $\mu^\mathrm{th}$-component of the discrete Fourier transform, which are defined in Appendix~\ref{sec:appA}. 

Modes of each resonator with identical angular momentum are linearly coupled with the inter-resonator coupling rate $\Jc_\mu$. In contrast, the Kerr nonlinearity couples all the modes within each resonator via four-wave mixing (FWM) processes. The interplay between the linear coupling in the spatial dimension and the nonlinear coupling in the frequency dimension is the source of the rich dynamics of the system. 

In the present section, we restrict ourselves to a linear and single mode analysis by considering only the central mode $\mu=0$ with $\gc=0$. The critical coupling conditions are of particular interest for maximizing the pump transfer to the resonators and the eigenvalue analysis of the coupled mode matrix for finding the EP conditions, which separate two conceptually different dimer states. 
\subsection{Critical coupling conditions}\label{sect:crit_couple}
Let us analyse the condition for critical coupling. In the linear single-mode representation, Eq.~\ref{eq:cLLEs} is simplified as follows: 
\begin{align}
i\frac{\mathrm{d}}{\mathrm{d}t}\begin{pmatrix}A\\B\end{pmatrix} &=
M\begin{pmatrix}A\\B\end{pmatrix}
+\sqrt{\kappa_\mathrm{ex,1}} \begin{pmatrix}s_\mathrm{in}\\0\end{pmatrix}, \notag\\
M & = \begin{pmatrix}
\tfrac{1}{2}\delta  - \tfrac{i}{4}\Delta\kappa_\mathrm{ex} & -\Jc\\
-\Jc & -\tfrac{1}{2}\delta  + \tfrac{i}{4}\Delta\kappa_\mathrm{ex}
\end{pmatrix}\notag\\
&\qquad + (\omega_0 - \omega_p  - \tfrac{i}{2}\kappa_\mathrm{ex}) I, \notag\\
s_\mathrm{out,1} &= s_\mathrm{in}-\sqrt{\kappa_\mathrm{ex,1}}A,\notag\\
s_\mathrm{out,2} &= \sqrt{\kappa_\mathrm{ex,2}}B.
\label{eq:2resLin}
\end{align}
In Eq.~\ref{eq:2resLin}, we defined the external coupling mismatch $\Delta\kappa_\mathrm{ex} = \kap1 -\kap2$ and the average external coupling $\kappa_\mathrm{ex} = \frac{1}{2}(\kap1+\kap2)$. The identity matrix is denoted as $I$. Critical coupling is achieved when the transmission via the through port [Fig.~\ref{fig:criticalCouplingSimple}(a)] vanishes, i.e. $s_\mathrm{out,1}=0$. In the case of a single resonator, critical coupling is achieved when the external coupling rate matches the loss, i.e. $\kappa_{\mathrm{ex}} = \kappa_0$~\cite{yariv2002critical,matsko2006optical}. For two resonators, the conditions are easily found in case of $\delta=0$. There are two possibilities
\begin{align}
\kappa_{\mathrm{ex},1}/\kappa_0  &= \frac{4(\Jc/\kappa_0)^2+\kappa_{\mathrm{ex},2}/\kappa_0+1}{\kappa_{\mathrm{ex},2}/\kappa_0+1}\label{eq:critCoupl1}\\
\kappa_{\mathrm{ex},1}/\kappa_0 & = 2 + \kappa_{\mathrm{ex},2}/\kappa_0.
\label{eq:critCoupl2}
\end{align} 
Eq.~\ref{eq:critCoupl1} is a natural generalization of the critical coupling conditions for a single resonator that can be achieved by setting $\Jc$ to zero. Eq.~\ref{eq:critCoupl2} satisfies the critical coupling condition at $\omega_p = \omega_0 \pm \sqrt{4 \Jc^2 - (\kappa_0+\kappa_{\mathrm{ex},2})^2}$. This condition requires strong coupling, i.e. $\Jc>\frac{1}{2}(\kappa_0+ \kappa_{\mathrm{ex},2})$. The critical coupling conditions are shown in Fig.~\ref{fig:criticalCouplingSimple}(d) for $\kappa_{\mathrm{ex},2} =0 $. Typical cavity field intensities for both cases are plotted in Fig.~\ref{fig:criticalCouplingSimple}(b,c) as a function laser detuning. 

The first critical coupling condition given by Eq.~\ref{eq:critCoupl1} has a quadratic dependence on the inter-resonator coupling rate [Fig.~\ref{fig:criticalCouplingSimple}(d)]. It leads to a broad resonance with a dip in the first resonator (blue) and a narrow resonance in the second resonator (red) at the same resonance frequency [Fig.~\ref{fig:criticalCouplingSimple}(c)]. The second critical coupling condition (Eq.~\ref{eq:critCoupl2}) branches off the first one at $\Jc/\kappa_0 = 0.5$ and does not depend on the inter-resonator coupling rate. It features split resonances with identical linewidths [Fig.~\ref{fig:criticalCouplingSimple}(b)].

Experimental implementation of the multimode photonic dimer demonstrated the presence of the non-vanishing inter-resonator detuning $\delta$ caused by the fabrication imperfectness. Nonetheless, the possibility to control and manipulate $\delta$, and thereby establish control over the output solitonic spectrum, has been demonstrated and efficiently implemented by imprinting heaters directly on the photonic device~\cite{miller2015tunable,tikan2020emergent}. Critical coupling at non-vanishing $\delta$ is possible as well. The inter-resonator detuning introduces asymmetry in the distribution of the supermodes (eigenvectors) in each resonator. Thus, the supermode confined in the first (second) resonator requires smaller (larger) $\kap1$ to be critically coupled. It follows that in general when $\delta\neq 0$ only one supermode can be critically coupled for a given value of $\kap1$. For more details, see Appendix~\ref{sect:C}.

The qualitative behavior of the photonic dimer can be anticipated by examining the eigenvalues of the system Eq.~\ref{eq:2resLin}. Operating with a naturally Non-Hermitian system, we can exploit the concept of EP~\cite{Miri2019Exceptional} to shed light on the nature of each critical coupling conditions.
\subsection{Eigenvalues and exceptional points}
The eigenvalues of the matrix $M$ defined in Eq.~\ref{eq:2resLin} in case of $\delta=0$, $\kap2 = 0$, and $\omega_0 = \omega_p$ are given by 
\begin{align}
    \sigma_\pm = -i (\tfrac{1}{2}\kappa_0 + \tfrac{1}{4}\kap1) \pm \tfrac{1}{4}\sqrt{16 J^2-\kap1 ^2},
    \label{eq:eigenvalues}
\end{align} 
where the real (imaginary) part corresponds to resonance frequency (loss rate). The eigenvalues are shown in Fig.~\ref{fig:criticalCouplingSimple}(e) as a function of $\kap1$ for an inter-resonator coupling $J=1.5\kappa_0$. Two different regions of \emph{split resonance and split dissipation} are identified and shaded in Fig.~\ref{fig:criticalCouplingSimple} with orange and purple, respectively. For $\kap1 < 6\kappa_0$, the eigenvalues have degenerate imaginary part and split real parts, associated with the split resonances as depicted in Fig.~\ref{fig:criticalCouplingSimple}(b). In contrast, $\kap1>6\kappa_0$ leads to degenerate real parts and split imaginary parts, i.e. to identical resonance frequencies but different loss rates, as can be seen in Fig.~\ref{fig:criticalCouplingSimple}(c). The two regions correspond to the $\mathcal{PT}$-symmetric and $\mathcal{PT}$-symmetry broken states, respectively.

An EP is found between the two regions at $\kap1 = 6 \kappa_0$, where the system eigenvalues become degenerate and the two eigenvectors coalesce because of the vanishing square root in Eq.~\ref{eq:eigenvalues}. \emph{EPs lie along the line defined by $\kap1 = 4\Jc$, which separates the two critical coupling conditions in the $(\Jc,\kap1)$ plane.} Remarkably, the two critical coupling conditions and the EP line fork at $\Jc=\tfrac{1}{2}\kappa_0,~\kap1=2\kappa_0$. This particular point is highlighted by the blue star in Fig.~\ref{fig:criticalCouplingSimple}(d). It is the only EP that satisfies a critical coupling condition. This point also marks the entry into the strong coupling regime ($\Jc>\tfrac{1}{2}\kappa_0$). 
Above reasoning is valid when $\kap2 = 0$. In the general case ($\kap2 \neq 0$) the line of EPs can cross the line corresponding to critical coupling conditions.

Concluding, there are two types of critical coupling conditions in the photonic dimer. These conditions are found on both sides of the EPs, such that critical coupling can be achieved in the $\mathcal{PT}$-symmetric as well as $\mathcal{PT}$-symmetry broken states. In the next sections, we examine the versatile nonlinear dynamics and dissipative Kerr soliton generation in theses states. 





\section{\label{sec:3} Critically coupled resonators: split resonance (\texorpdfstring{$\mathcal{PT}$}{PT}-symmetric)}

In the present section we discuss the case of the split resonance ($\mathcal{PT}$-symmetric). We revisit ideas presented earlier in ~\cite{tikan2020emergent} by looking at the nonlinear dynamics from the supermode perspective. We demonstrate a separability of the GS dynamics from DWs living in the S supermodes. Finally, we show how this representation explains the origin of soliton hopping effect. An essential part of the investigation of the dynamics inherent to the photonic dimer and described by Eq.~\ref{eq:cLLEs} relies on numerical simulations.

\subsection{\label{sec:22} Four-wave mixing pathways between supermodes}

\begin{figure*}
	\centering
    \includegraphics[width=0.98\linewidth]{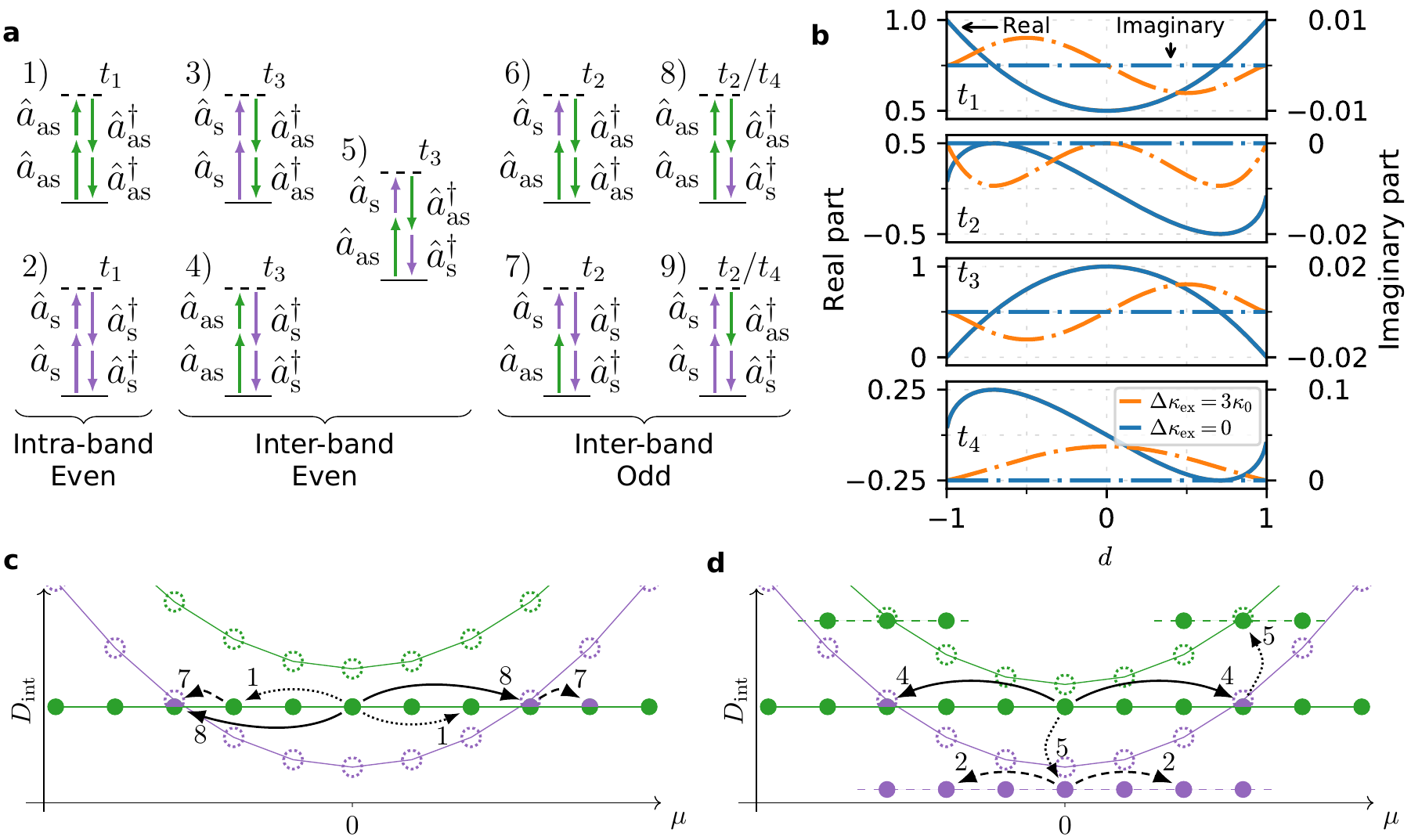}
	\caption{\textbf{FWM pathways between the dimer supermodes in split resonance ($\mathcal{PT}$-symmetric) regime. }(a) Table of the 9 possible FWM pathways represented with quantum operators and corresponding coefficients $t_i$. (b) Real (solid) and imaginary (dashed) part of nonlinear coupling coefficients as a function of the normalized inter-resonator detuning $d$ for 2 values of the external coupling mismatch $\Delta\kappa_\mathrm{ex}$ in the split resonance regime ($\Jc = 20 \kappa_0$). $\Delta\kappa_\mathrm{ex}$ is the source of the imaginary component, which however is small compared to the real part, except for $t_4$ around $d=0$. (c,d) Schematic integrated dispersion profile with emphasized FWM pathways. The labelling indexes are specified in (a). Empty dashed circles denotes cold cavity mode, filled circle denotes Kerr comb mode, and color codes AS (green) or S (purple) mode family. (c) GS generated in the AS supermodes via conventional FWM (even intra-band processes) and the emergence of commensurate dispersive waves via odd inter-band processes. (d) Soliton in the AS supermodes and the generation of a soliton in the S supermodes caused by even processes.}
	\label{fig:nonLinCoupl_process}
\end{figure*}

The linear part of Eq.~\ref{eq:cLLEs} can be diagonalized by a linear transformation on each pair of modes with index $\mu$. We define the complex inter-resonator detuning $\delta_c = \delta - i \tfrac{1}{2}\Delta\kappa_\mathrm{ex}$. If the inter-resonator coupling is independent of the wavelength, the complex frequency splitting $\Delta \omega_c = \sqrt{4\Jc^2+\delta_c^2}$, and the (complex) normalized inter-resonator detuning \[d_c \equiv \delta_c/\Delta\omega_c\] are independent of the mode index. Therefore, the non-unitary transformation diagonalizing the linear part of Eq.~\ref{eq:cLLEs} is given by $\ASmu = \alpha A_\mu + \beta B_\mu, ~\AASmu = \beta A_\mu - \alpha B_\mu$, where
\begin{align}
\alpha & \equiv \frac{\sqrt{1-d_c}}{\sqrt{2}} & \beta & \equiv \frac{\sqrt{1+d_c}}{\sqrt{2}}.
\end{align}	
Here S and AS stand for the symmetric (S) and antisymmetric (AS) mode, as they are completely symmetric (antisymmetric) at $d_c = 0$ in the split resonance regime. Then, by defining the spatial envelope of the field in the S and AS modes $\AS(\theta) = \sum_\mu \ASmu e^{i\mu\theta}, ~\AAS(\theta) = \sum_\mu \AASmu e^{i\mu\theta}$, we can express Eq.~\ref{eq:cLLEs} in the supermode basis (see details in Appendix~\ref{sec:appB} and alternative Hamiltonian formulation in~\cite{tikan2020emergent}):
\begin{subequations}
\label{eq:nlcLLEs}
\begin{align}
\frac{\mathrm{d}}{\mathrm{d}t} \ASmu =&[-i(\omega_\mu- \mu D_1-\omega_p-\tfrac{1}{2}\Re(\Delta \omega_c)) \notag\\
&\quad- \tfrac{1}{2}(\kappa_0 + \kappa_\mathrm{ex}+\Im(\Delta \omega_c))]\ASmu \notag\\
&\quad+\delta_{\mu,0}\alpha \sqrt{\kappa_{\mathrm{ex},1}}s_\mathrm{in} + i\gc \FFT\left[t_1 \AS\abssq{\AS} \right. \notag\\
 & \quad + t_2 \AAS\abssq{\AS} + t_4 \AS^2\AAS^* +t_3\AS \abssq{\AAS} \notag\\
 &\quad \left. + \tfrac{1}{2}t_3\AAS^2\AS^*- \tfrac{1}{2}t_2 \AAS\abssq{\AAS}\right]_\mu\label{eq:nlcLLEsA}\\ 
\frac{\mathrm{d}}{\mathrm{d}t} \AASmu =& [-i(\omega_\mu- \mu D_1 -\omega_p+\tfrac{1}{2}\Re(\Delta \omega_c))\notag\\
&\quad- \tfrac{1}{2}(\kappa_0 + \kappa_\mathrm{ex}-\Im(\Delta \omega_c))] \AASmu \notag\\
&\quad +\delta_{\mu,0}\beta \sqrt{\kappa_{\mathrm{ex},1}}s_\mathrm{in} + i\gc	\FFT\left[ \tfrac{1}{2}t_2 \AS\abssq{\AS} \right. \notag\\
& \quad + t_3\AAS\abssq{\AS} + \tfrac{1}{2}t_3 \AS^2\AAS^* -t_2 \AS \abssq{\AAS}\notag\\
&\quad \left. - t_4 \AAS^2\AS^*+t_1\AAS\abssq{\AAS} \right]_\mu.
\label{eq:nlcLLEsB}
\end{align}
\end{subequations}
As one can see, the linear anti-diagonal terms  are eliminated in the supermode basis, while the nonlinear terms (diagonal in the resonator basis) induce nonlinear coupling between the supermodes.  In particular, we identify FWM processes between the supermodes. The efficiencies of FWM pathways are associated with nonlinear coupling coefficients $t_i,~i=1,..,4$, defined in Eq.~\ref{eq:def_nccs} of Appendix~\ref{sec:appB}. For example, the term $\tfrac{1}{2}t_3\AAS^2\AS^*$ in Eq.~\ref{eq:nlcLLEsA} signifies the annihilation of two photons in the AS supermode and the creation of two photons in the S supermode. The rate of the process is proportional to $\gc t_3$. Each FWM process can be represented by a quantum Hamiltonian term, for example
\begin{equation}
\opa_{\sym,\mu_1}^\dagger\opa_{\sym,\mu_2}^\dagger\opa_{\ant,\mu_3}^{}\opa_{\ant,\mu_4}^{},
\end{equation} 
where $\opa_{i,\mu}^\dagger$ ($\opa_{i,\mu}^{}$) is the  annihilation (creation) operator for supermode $i=\sym,~\ant$ with longitudinal mode index $\mu$, and $\mu_1 + \mu_2 = \mu_3 + \mu_4$.
The 9 different non-linear processes from Eq.~\ref{eq:nlcLLEs} are depicted in Fig.~\ref{fig:nonLinCoupl_process}(a), where they are arranged in categories corresponding to \emph{intra-band} even processes, \emph{inter-band} even processes and inter-band odd processes. 

We refer to a nonlinear process as \emph{intra-band} when two annihilated and two created photons  are from the same supermode family, while inter-band processes imply nonlinear mixing of photons belonging to different supermodes, inspired by the concept of Bloch bands in condensed matter Physics.
The number parity of the process (even or odd) refers to the number of photons from each supermode family that is involved.
We note that processes (2,4,7,9) are the counterparts of processes (1,3,6,8) for permuted supermodes index. Schemes of possible FWM pathways between the supermodes (while a solitonic state is generated in the AS supermode family) are shown in Fig.~\ref{fig:nonLinCoupl_process}(c,d) 
These processes are distinguished by the nature of FWM: Fig.~\ref{fig:nonLinCoupl_process}(c) shows odd processes (except the conventional even process $\#$1, associated with Hamiltonian term $\opa_{\ant,\mu_1}^\dagger\opa_{\ant,\mu_2}^\dagger\opa_{\ant,\mu_3}^{}\opa_{\ant,\mu_4}^{}$), while Fig.~\ref{fig:nonLinCoupl_process}(d) shows even processes leading to soliton hopping (see Sec.~\ref{sec:solHopping}). 

While the index $\mu$ has been omitted in Fig.~\ref{fig:nonLinCoupl_process}(a) for readability, both the mode number and the energy have to be conserved in a FWM process. We employ the concept of integrated dispersion $D_\mathrm{int}(\mu) = \omega_\mu - (\omega_0 + D_1\mu)$ to depict the processes which satisfy the phase matching conditions.


The real and imaginary parts of the nonlinear coupling coefficients are shown in Fig.~\ref{fig:nonLinCoupl_process}(b) as a function of the normalized inter-resonator detuning $d=\delta/\sqrt{4J^2+\delta^2}$ with solid and dashed dotted lines, respectively. The parameters are chosen in the split resonance regime with $J=20\kappa_0$. Vanishing and non-vanishing $\Delta\kappa_\mathrm{ex}$ are considered, emphasizing that the imaginary parts of all the nonlinear coupling coefficients originate from the external coupling mismatch. The imaginary part, however, generally constitutes only a small fraction of the absolute value of the nonlinear coupling coefficients.

Coefficient $t_1$ is responsible for the \emph{intra-band} processes, that is, the usual FWM within the same mode family (S) or (AS). It has its lowest value equal to 0.5 at the maximum hybridization ($d=0$). In contrast, coefficient $t_3$ is \emph{maximized} at $d=0$ and causes \emph{inter-band and even} processes. The coefficients $t_2$ and $t_4$ are responsible for \emph{inter-band} and odd processes. Their real parts are odd with respect to $d$. Therefore, there are \emph{no odd FWM processes} at $d=0$, unless an external coupling mismatch is present. In this case, coefficient $t_4$ has a non-vanishing absolute value.


\subsection{Nonlinear dynamics and soliton generation in split resonance (\texorpdfstring{$\mathcal{PT}$}{PT}-symmetric) regime}
\begin{figure*}
	\centering
	\includegraphics[width=0.75\linewidth]{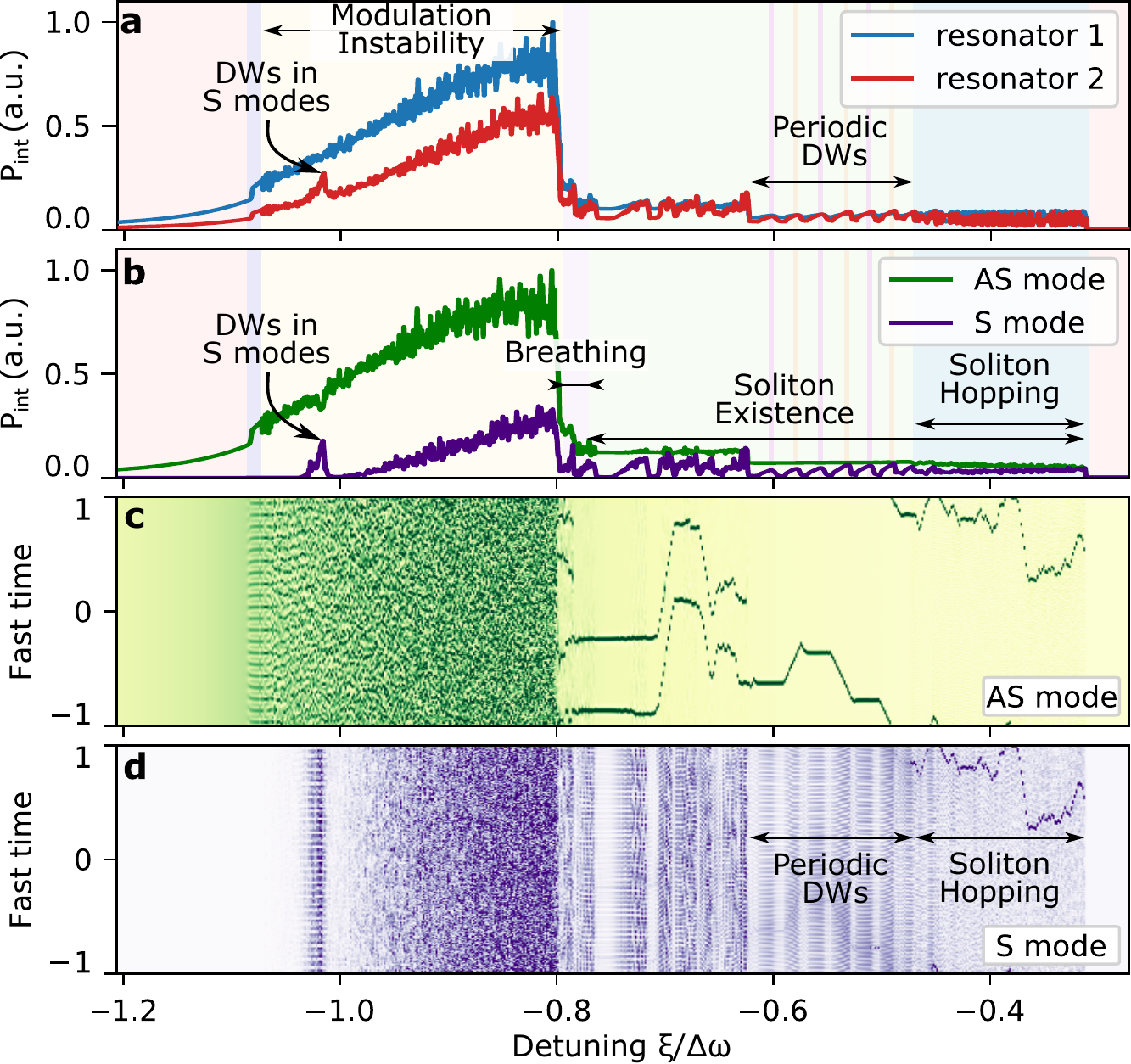}
	\caption{\textbf{Numerical simulations of the split resonance regime ($\mathcal{PT}$-symmetric case).} (a) Intracavity field power evolution in resonator basis. (b) Intracavity field power evolution in supermode basis. (c,d) Spatiotemporal diagrams for the field evolution in AS and S supermodes, respectively. Parameters of numerical simulations are close to the exact critical coupling conditions: $\kappa_0/2\pi$ = 50 MHz, $\kappa_{\mathrm{ex},1}/2\pi$ = 100 MHz, $\kappa_{\mathrm{ex},2}/2\pi$ = 20 MHz, $\delta/2\pi$ = 4 GHz, J$/2\pi$ = 4.5 GHz, the pump power was fixed to 1.2 W.}
	\label{fig:sim_cc_2}
\end{figure*}

The type of the critical coupling conditions corresponding to the split resonances allows for accessing dynamical states characterized by efficient generation of bright dissipative Kerr solitons in both cavities. Emergent dynamical effects described in~\cite{tikan2020emergent} are found in this regime. In this section, we provide an additional (to the result already shown in~\cite{tikan2020emergent}) and complementary description of these phenomena by representing the inter-resonator field in the hybridized supermodes basis. 

\subsubsection{\label{sec:321} Modulation Instability state}
We restrict ourselves to the AS supermodes pumping scheme since we did not observe dynamics different from the single resonator case when exciting the S supermode family. Fig.~\ref{fig:sim_cc_2}(a,b) show the intracavity power evolution as a function of laser detuning in the resonator and supermode basis, respectively. It is numerically generated by exciting the system in a soft manner, i.e. adiabatically changing the laser detuning $\xi = \omega_0 -\omega_p$ from blue to red side of the AS hybridized resonance. Initial dynamics is found to be similar to the single resonator case. We observe the formation of primary combs in the AS supermode family followed by cnoidal waves (Turing rolls). The subsequent chaotic modulation instability stage [yellow area in Fig.~\ref{fig:sim_cc_2}(a,b)] already demonstrates a significant difference. Namely, the average intracavity power evolution in the second resonator, which is depicted by the red line, as a function of normalized detuning $\xi/\Delta\omega$ \emph{exhibits a local maximum} inside the modulation instability area which corresponds to the efficient photon transfer to the S supermodes [violet curve in Fig.~\ref{fig:sim_cc_2}(b)]. At these values of detuning ( $\xi/\Delta\omega \approx$ -1), we observe an \emph{enhancement of spectral components} distinct from the modulation instability gain region.
The mode number of the components correspond exactly to the distance from the pumped mode to the lower (S supermodes) parabola for a given value of the laser detuning, as described in~\cite{tikan2020emergent}. 
This is a first signature of the interaction between the supermodes. 

Fig.~\ref{fig:sim_cc_2}(c,d) provide the underlying evolution of intracavity power (spatiotemporal diagram) 
in the supermode basis. The modulation instability region in the conventional basis does not differ for the single-particle dynamics. However, the supermode basis reveals that the transfer of photons to the S supermode family occurs after a certain detuning threshold. As follows from the spatiotemporal diagram of the AS state, it occurs in the developed AS supermodes modulation instability stage, where collision and annihilation of unstable coherent structures lead to the enhancement of wings in the optical spectrum~\cite{Coulibaly2019Turbulence} and thereby populates the modes in vicinity of the symmetric resonances.

\subsubsection{\label{sec:322} Breathing state}

\begin{figure*}
	\centering
	\includegraphics[width=\linewidth]{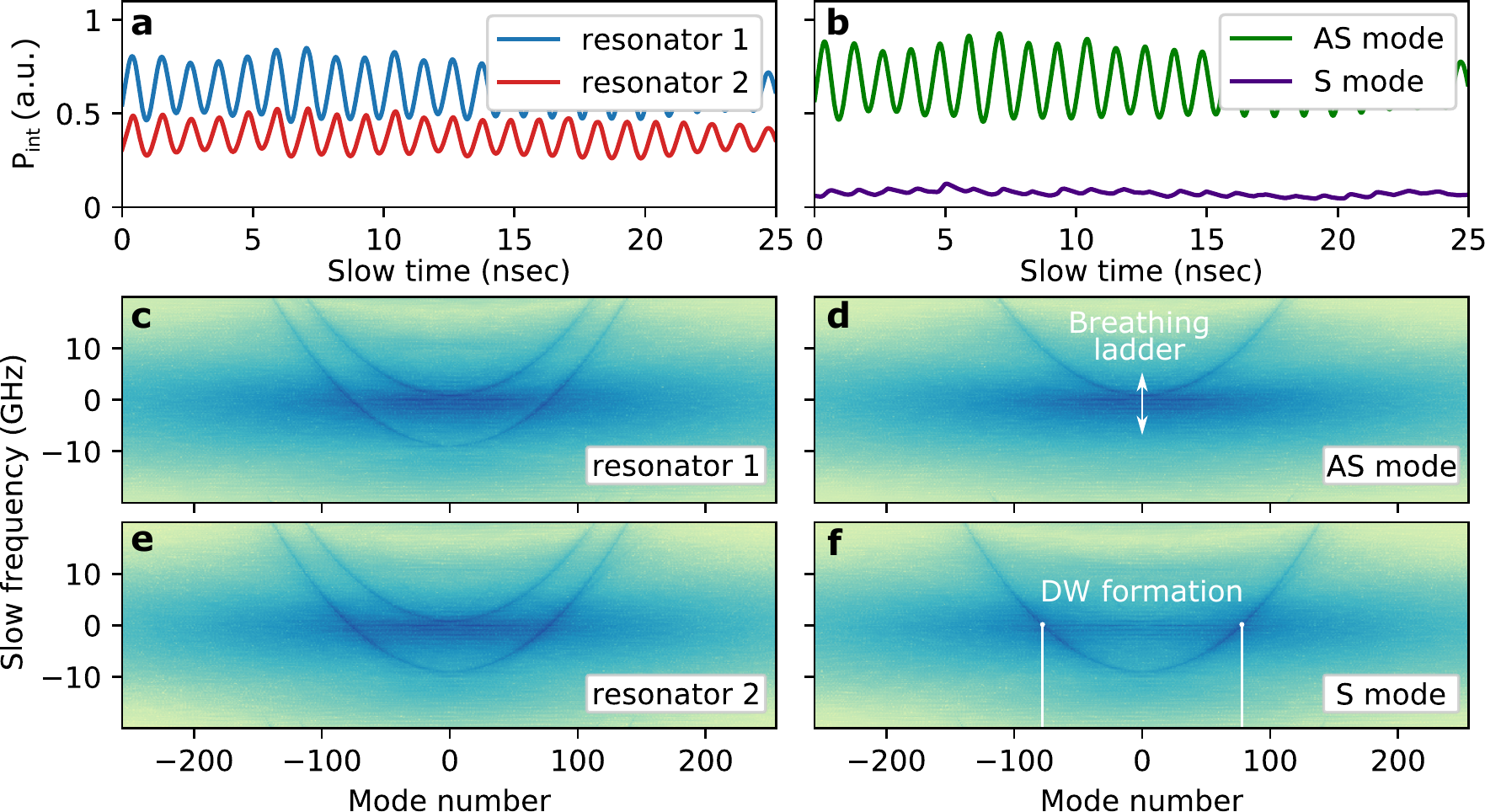}
	\caption{\textbf{Breathing state.} (a,b) Intracavity power evolution with a fixed pump laser detuning (a) in the first and second resonator (shown by blue and red lines) and (b) in the supermode basis (green and violet for AS and S mode families, respectively). (c-f) Nonlinear dispersion relations (c,e) for the first and second resonators (d,f) for AS and S supermode families.}
	\label{fig:sim_cc_2_breathers}
\end{figure*}

As in conventional single resonator systems above a threshold pump power level, the modulation instability region is followed by the breathing solitons region [violet area in Fig.~\ref{fig:sim_cc_2}(a,b)]. Breathing originates from the Hopf bifurcation as demonstrated for the single resonator case~\cite{Leo2013Dynamics}. It manifests itself as a periodic oscillation of localized coherent structures (similar to solitons on a finite background, such as Kuznetsov-Ma soliton~\cite{kuznetsov1977solitons,kibler2012observation} - a solution of the nonlinear Schr\"odinger equation), which radiates DWs at every cycle of oscillation. Fig.~\ref{fig:sim_cc_2_breathers}(a) shows the intracavity power evolution. Breathers in the photonic dimer exist in both resonators and oscillate \textit{in phase}. However, the intracavity trace is found to be randomly deviating from the average because of the photon transfer to the S supermodes and, therefore, the generation of additional DWs.

The periodic oscillation of a coherent structure in slow time results in the appearance of a ladder of straight and equally-spaced lines on the nonlinear dispersion relation~\cite{Tusnin2020Nonlinear}. The presence of such ladder has been experimentally demonstrated (see~\cite{tikan2020emergent} supplementary information) by reconstructing the comb spectrum with high resolution. Indeed, as follows from  Fig.~\ref{fig:sim_cc_2_breathers}(c,e) the breathing frequency is given by the frequency offset between the lines. The same reasoning can be applied to the single resonator breathing states. There, it has been demonstrated experimentally that the breathing frequency linearly depends on the pump laser detuning~\cite{Lucas2017Breathing}. Therefore, we can make a conjecture that the breathing occurs due to the photon transfer between the Kerr-shifted dispersion parabola and the first solitonic line given by the laser detuning, while the breathing frequency is the corresponding gap.
In the points where the ladder crosses the AS supermodes parabola enhancement of the comb power is observed. Therefore, optical spectrum of a breather contains a set of sidebands~\cite{Lucas2017Breathing}.

In the supermode basis [see Fig.~\ref{fig:sim_cc_2_breathers}(b)] it becomes evident that the breathing occurs mostly in AS supermode families. 
Therefore, the breathing dynamics in the AS mode family does not show significant difference from the conventional breathing found in the single resonator case as follows from the nonlinear dispersion relation [Fig.~\ref{fig:sim_cc_2_breathers}(d)], although it demonstrates significant differences in the resonator basis.
Fig.~\ref{fig:sim_cc_2_breathers}(f) shows the nonlinear dispersion relation for the S supermode family. The origin of the DWs which perturb the breathing state can be seen as an enhancement of the certain supermodes in the S family ($\mu\approx\pm 70$) in the places where the ladder from AS supermodes crosses the S parabola.

\subsubsection{Soliton hopping state\label{sec:solHopping}}

\begin{figure}
	\centering
	\includegraphics[width=\linewidth]{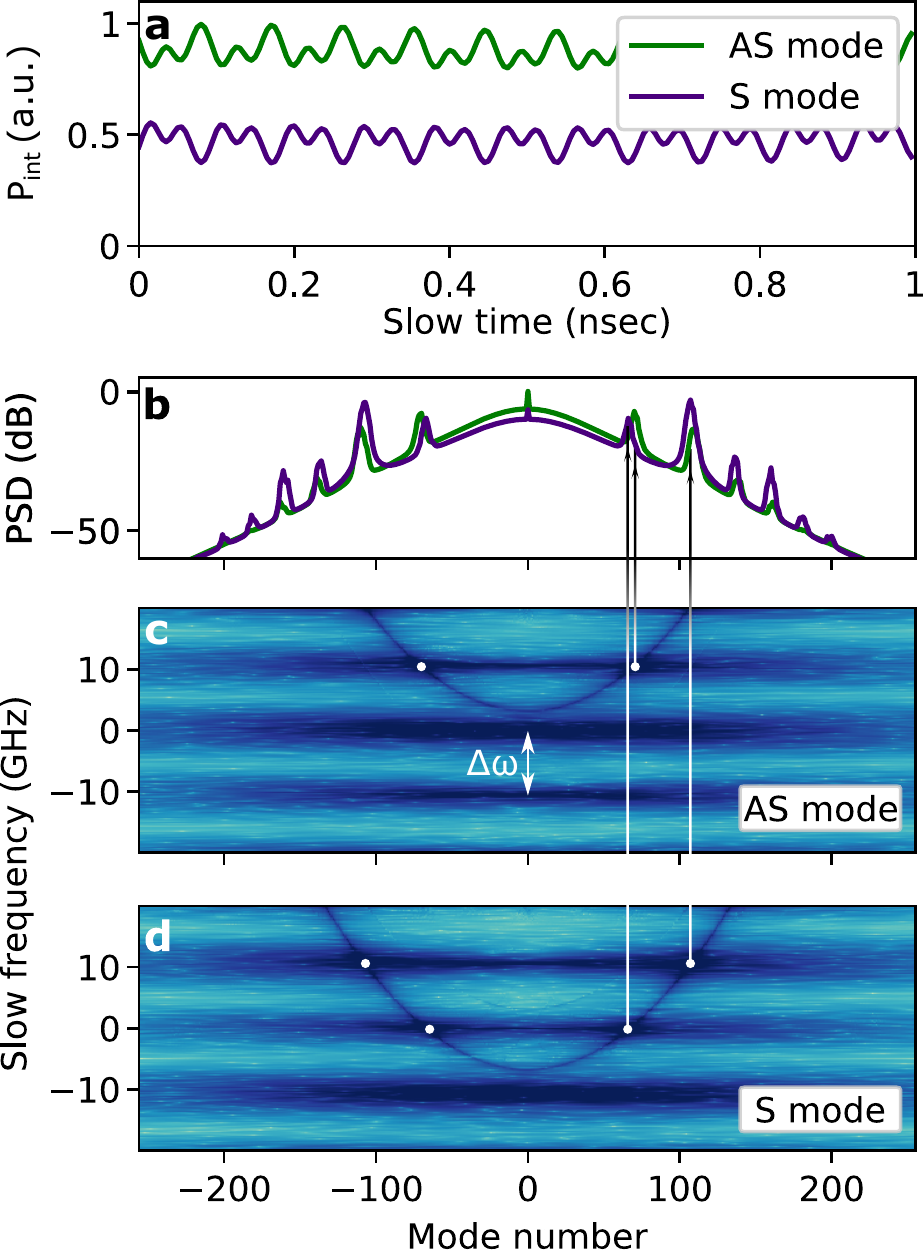}
	\caption{\textbf{Soliton hopping state.}
(a,b) Intracavity field power evolution and power spectral density (PSD) at fixed laser detuning corresponding to the the soliton hopping regime, in supermode basis.
(c,d) Nonlinear dispersion relation for AS and S supermode families, respectively. White dots show points on the nonlinear dispersion relation where the spectral components are enhanced. Arrows connect these points with the corresponding spectral components.}
	\label{fig:sim_cc_sh}
\end{figure}

The soliton hopping state recently predicted in the photonic dimer~\cite{tikan2020emergent} is characterized by a periodic energy exchange between the coupled resonators in the presence of temporally-localized coherent structures. Inter-resonator oscillations have a frequency equal to the splitting between the supermode parabolas. The average power modulation is much stronger than in the breathing state, which leads to the enhanced sideband amplitudes in the optical spectrum [see Fig.~\ref{fig:sim_cc_sh}(b)].

The spatiotemporal diagrams for the diagonalized system [see Fig.~\ref{fig:sim_cc_2}(c,d)] reveal a remarkable insight about the soliton hopping state. The soliton hopping range [blue area in Fig.~\ref{fig:sim_cc_2}(a,b)] coincides with the emergence of a localised coherent structure in the S supermodes family accompanied by a characteristic solitonic step in the average intracavity power evolution as follow from Fig.~\ref{fig:sim_cc_2}(b). This coherent structure is generated via the emerged FWM pathways depicted in Fig.~\ref{fig:nonLinCoupl_process}(d). Soliton in AS supermode family acts in this case as a source of photons which triggers the parametric processes, thereby resonantly populating the S parabola in the vicinity of the 0$^\mathrm{th}$ mode (i.e. with the offset $-\Delta\omega$) via the process $\#$5 ($\opa_{\ant,\mu_1}^\dagger\opa_{\sym,\mu_2}^\dagger\opa_{\ant,\mu_3}^{}\opa_{\sym,\mu_4}^{}$). Energy conservation is ensured by populating supermodes offset by approximately $+\Delta\omega$. Cascaded parametric process $\# 2$ ($\opa_{\sym,\mu_1}^\dagger\opa_{\sym,\mu_2}^\dagger\opa_{\sym,\mu_3}^{}\opa_{\sym,\mu_4}^{}$) populates the neighbouring S supermodes similarly to the CW-pumped single resonator. Therefore, we assume that the coherent structure generated in S supermodes is a GS (i.e. supermode dissipative Kerr soliton). Thus, the origin of the oscillatory behaviour can be seen as a time periodic interference of coherent structures living in different supermodes.

Fig.~\ref{fig:sim_cc_sh}(a) shows the dimer dynamics at fixed pump laser detuning, in the soliton hopping regime in AS and S supermodes representation. It can be obtained numerically by seeding the solitonic state in the AS supermode (see Appendix~\ref{sec:ApD}) and further tune into the soliton hopping state.  The average power exhibits small amplitude oscillations around a certain value. Periodic oscillations in slow time results into the series of sidebands (similar to Kelly-sidebands widely present in the mode-locked lasers~\cite{kelly1992characteristic}) in the optical spectrum as has been shown in~\cite{tikan2020emergent}. Corresponding nonlinear dispersion relation shows a ladder of lines similar to the breathing state discussed in Sec.~\ref{sec:322}, but the spacing between them is equal to the splitting between the DWs parabolas. The origin of the double maxima spectral sidebands is well seen in the supermode basis [see Fig.~\ref{fig:sim_cc_sh}(b)]. They appear due to the different Kerr nonlinearity-induced shift of supermodes in the presence of inter-resonator detuning. White dots indicate the point where the ladder crosses dispersive parabolas and the continuing arrows indicate the corresponding spectral components enhancement. Both nonlinear dispersion relations depicted in Fig.~\ref{fig:sim_cc_sh}(d,c) contain a corresponding DW parabola and the hopping ladder. Since the ladder crosses parabolas at slightly different mode numbers, sidebands have two maxima.


\begin{figure*}
	\centering
	\includegraphics[width=0.8\linewidth]{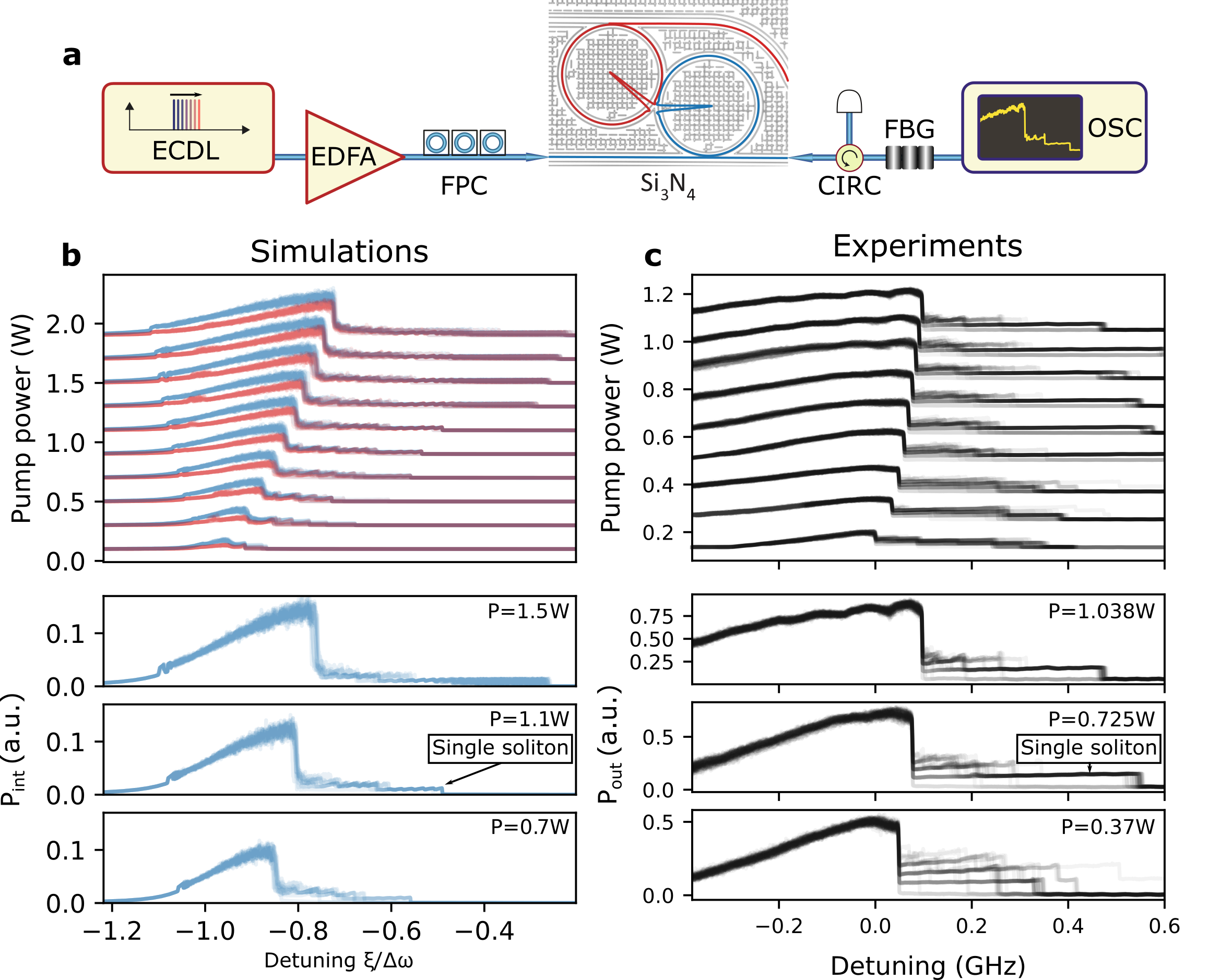}
	\caption{\textbf{Experimental and numerical evidences of single GS generation.} (a) Experimental setup for a single GS generation with an integrated Si$_3$N$_4$ photonic dimer driven by external cavity diode laser (ECDL). CW light is amplified by an erbium doped fiber amplifier (EDFA) and after passing a fiber polarization controller (FPC) is injected in the photonic dimer. The generated light in filtered by a fiber Bragg grating (FBG) and recorded with a fast oscilloscope (OSC). (b,c) Numerical and experimental confirmations of the single GS generation, respectively. (b, top) Simulation of average intracavity field evolution for different input powers as a function of pump laser detuning. For every value of the pump power there are 10 traces superimposed.  (b, bottom) Examples at three different powers. Single GSs are depicted by arrow. (c) Experimental confirmation obtained by piezo-tuning of the central frequency of a widely tunable ECDL over the AS resonance. For every value of the pump power there are 50 generated light traces superimposed. }
	\label{fig:sing_gs_gen}
\end{figure*}

\subsection{\label{sec:43} Experimental and numerical evidences of the deterministic single soliton generation}

Deterministic generation of a single soliton state in optical microresonators is essential for a turn-key dissipative Kerr soliton-based broadband frequency combs generation. Indeed, passing the chaotic modulation instability stage, soliton arrangement inside the cavity can be arbitrary which leads to a non-homogeneous spectral profile due the interference of different solitonic components. One way to control and structure the soliton arrangement inside the cavity is to introduce a background modulation which leads to the generation of perfect soliton crystals~\cite{karpov2019dynamics}. However, the single soliton state is, nonetheless, difficult to achieve in this configuration.

Another way to naturally fall into the single soliton state has been described in~\cite{Bao2017Spatial}. It has been proposed to employ a strong avoided mode crossing with higher-order modes of the resonator, which leads to an extensive cavity Cerenkov radiation~\cite{brasch2016photonic}. In this case, the soliton, being a line on the nonlinear dispersion relation~\cite{Leisman2019EDR}, crosses the distorted cavity mode, which leads to effective photon transfer towards the higher-order modes~\cite{Yi2017Single}. In this way, every soliton  acts as a "source" of DWs. Therefore, if the energy of the DWs is sufficient to perturb the solitonic states the number of solitons will decay towards unity, where the state will be stabilized. 

Here, we present a deterministic version of this mechanism utilizing discovered properties of the photonic dimer~\cite{tikan2020emergent}. Due to the more complex dispersion landscape, the single soliton generation process does not require any additional interaction with higher-order mode, even though it is shown to be enhanced for certain supermodes due to the underlying symmetry~\cite{tikan2020symmetry}. Indeed, the periodic intra-resonator field enhancement due to the crossing of the S supermode parabola is found to be sufficient to trigger the process discussed in~\cite{Bao2017Spatial}. In order to verify this claim, we investigate both numerically and experimentally the GSs generation. Fig.~\ref{fig:sing_gs_gen} shows the qualitative comparison of numerical and experimental phase diagrams. As follows from the numerical simulations of coupled LLEs~\ref{eq:cLLEs}, single soliton generation occurs when passing a threshold power of 0.9 W. A similar result follows from the experimental investigations. A schematic of the experimental setup is shown in Fig.~\ref{fig:sing_gs_gen}(a). Single GSs are generated with an integrated Si$_3$N$_4$ photonic dimer driven by external cavity diode laser. The CW pump is amplified by an erbium doped fiber amplifier to achieve the power level need for the investigation of the threshold of the process. After passing a fiber polarization controller needed to guarantee that the solitons are generated in a single polarization mode family, the light is injected in the photonic dimer. The generated light in filtered by a fiber Bragg grating and recorded with a fast oscilloscope. Figure~\ref{fig:sing_gs_gen}(c) shows 50 traces of the generated combs power at different values of the pump power as a function of the laser detuning from the position of AS resonance. The central frequency of a widely tunable external cavity diode laser has been controlled by the piezo-tuning technique. Other details of the experimental measurements can be found in~\cite{tikan2020emergent}, Methods section.

\section{\label{sec:31} Critically coupled resonators: split dissipation (\texorpdfstring{$\mathcal{PT}$}{PT}-broken)}
Passing through an EP, which exhibits a singularity of nonlinear interactions efficiency, we enter the domain of split dissipation ($\mathcal{PT}$-symmetry broken phase), which exhibits drastically different dynamical features. We study soliton generation in this region and show that $\mathcal{PT}$-symmetry breaking leads to soliton generation in either cavities. The soliton localization can be switched by increasing the pump power and thereby flipping the broken $\mathcal{PT}$-symmetry. Four distinct dynamical states are identified, we observe among them \emph{on-demand} perfect soliton crystals generation, which can be a promising alternative to the existing technology relying on the resonator's mode interaction~\cite{karpov2019dynamics}.

\subsection{Nonlinear coupling coefficients and their divergence at the exceptional point}
The system of Eq.~\ref{eq:cLLEs} can be diagonalized to Eq.~\ref{eq:nlcLLEs} in the $\mathcal{PT}$-symmetry broken state in the same way as in the $\mathcal{PT}$-symmetric state, by applying a transformation matrix $T_m$ such that $T_m M T_m^{-1}$ is diagonal (see Eq.~\ref{eq:2resLin}). The diagonalization enables the computation of the nonlinear coupling coefficients describing the nonlinear interaction between the supermodes. The non-vanishing values of the nonlinear coupling coefficients are displayed in Fig.~\ref{fig:phase_space_case1}(c) as a function of $\Jc$, where the external coupling $\kap1$ is varied quadratically with $J$ to satisfy the critical coupling condition (Eq.~\ref{eq:critCoupl1}).
The nonlinear coupling coefficients are normalized by a real factor 
\begin{equation}
N \equiv \frac{i\Delta\kappa_\mathrm{ex}}{2\Delta\omega_c},
\end{equation} equal to the square of the norm of $T_m (1,0)^T$. Indeed, the system is non-Hermitian such that $T_m$ does not preserve the norm (the transformation is not Unitary). This means the variables $\abssq{\ASmu},~\abssq{\AASmu}$ \emph{are not proportional} to the number of photons. The normalization allows the interpretation of $t_i \gc/N$ as a rate per photon.

The nonlinear coupling coefficients behave differently in the regime of split dissipation with $d=0$ than in the regime of split resonance: $t_3$, associated with even inter-band processes, vanishes completely, while $t_2,~t_4$, ($t_1$) become purely imaginary (real). These are related to odd inter-band (even intra-band)  processes. At the EP ($J/\kappa_0 = 0.5$) they \emph{diverge}. 
The normalization corrects the divergence of  $t_1$ and $t_2$, but not of $t_4$, which exhibits a singularity at the EP.  
The constant value $t_1/N=1$ signifies that the rate of the intra-band FWM does not vary with $J$ in this critical coupling condition.

Although the linear concepts of $\mathcal{PT}$ symmetry are useful to understand the dynamics of the $\mathcal{PT}$-symmetry broken state, we observe that the solitonic states presented in the next section are not distributed spatially according to the supermode basis like in the $\mathcal{PT}$-symmetric regime. On the contrary, they are distributed in the resonator basis, which is diagonal with respect to nonlinearity.

\subsection{\label{sec:314} Phase diagram: inter-resonator coupling vs pump power}
\begin{figure*}
	\centering
	\includegraphics[width=\linewidth]{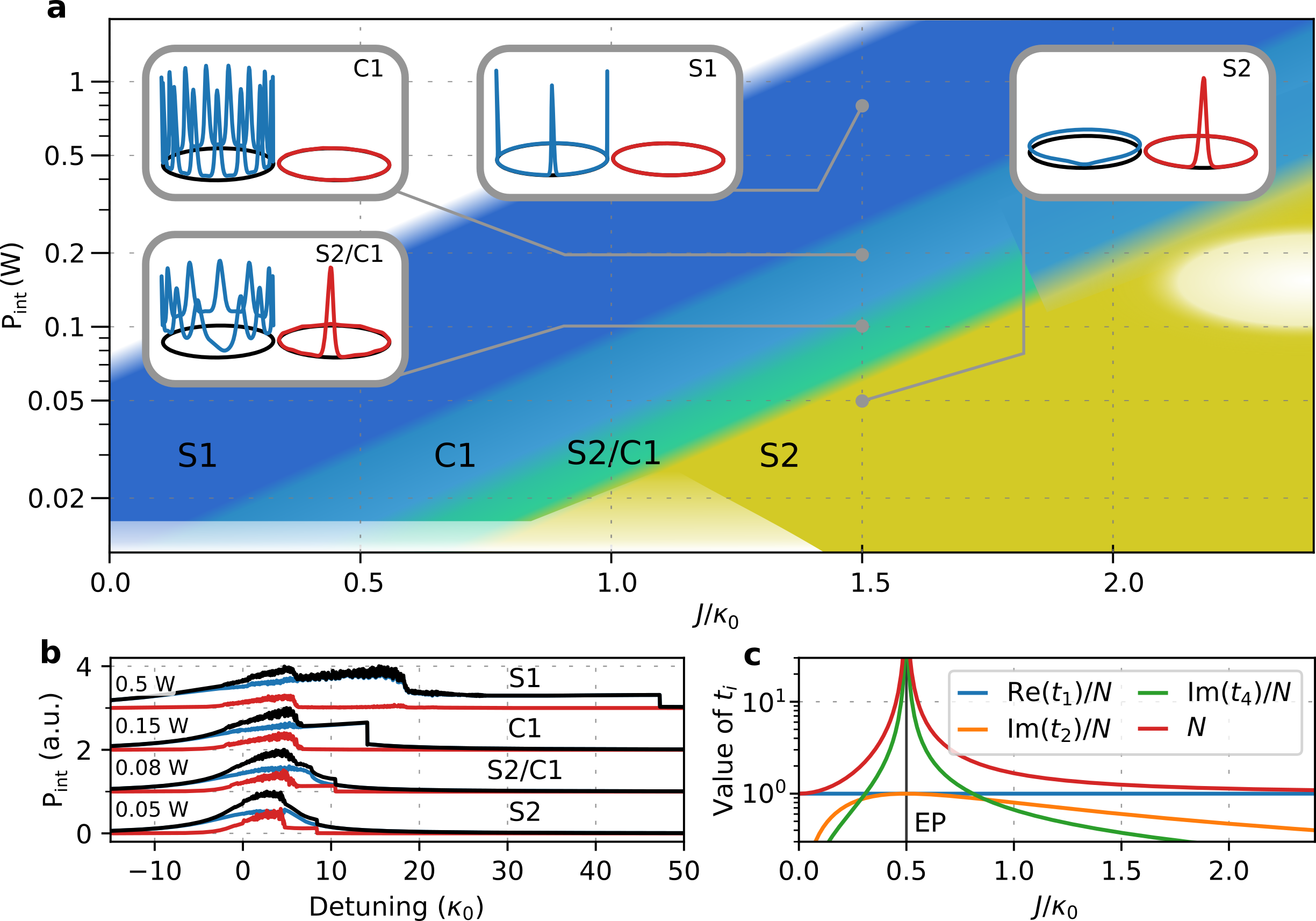}
	\caption{\textbf{Phase diagram for critically coupled photonic dimer under the split dissipation ($\mathcal{PT}$-broken) condition.} (a) Schematic phase diagram represented as a function of input power and $J$. Four states are identified: soliton in resonator 1 (S1, blue region), soliton crystal in resonator 1 (C1, cyan region), coexistence of soliton in resonator 2 and periodic coherent structure resonator 1 (S2/C1, green region), and soliton in resonator 2 (S2, yellow region). The insets show the intracavity field intensity in the resonators 1 (blue) and 2 (red), at the parameter indicated by the grey dot. White color refers to the absence of stable solitonic state. (b) Intracavity power traces at different input powers for $J/2\pi$ = 70 MHz. (c) Nonvanishing values of the nonlinear coupling coefficients and normalizing factor $N$ presented as a function of $\Jc$ along the critical coupling condition. The crossing of the EP at $\Jc/\kappa_0 = 0.5$ corresponds to a singularity in coefficients $t_1$, $t_2$, and $t_4$. Parameters for the simulations are $\kappa_0/2\pi=50$~MHz, $\kap2= 0$, $\delta=0$, $D_1/2\pi=$180~GHz, $D_2/2\pi=4$~MHz.}
	\label{fig:phase_space_case1}
\end{figure*}

We numerically explore the phase diagram under the condition of critical coupling in the nondegenerate dissipation regime [see Fig.~\ref{fig:phase_space_case1}(a)]. 
In Fig.~\ref{fig:phase_space_case1}(c), $\kap1$ is varied with $J$ in the way that the critical coupling condition (Eq.~\ref{eq:critCoupl1}) is satisfied across the phase diagram and the dimer is in a state of broken $\mathcal{PT}$-symmetry (split dissipation). An EP is found at $J=0.5\kappa_0$ [see Fig.~\ref{fig:criticalCouplingSimple}(d)]. We selected 14 values of pump power distributed logarithmically from 0.01~W to 1.5~W and 13 values of inter-resonator detuning distributed linearly from $0\kappa_0$ to $2.4\kappa_0$. For each set of parameters, we employ the conventional soliton generation scheme by scanning the resonance from blue to red-detuned side. The spatiotemporal and spectrum evolution diagrams in the resonator basis is used to identify the stable soliton state that is generated during the scan. Colors in Fig~\ref{fig:phase_space_case1}(a) correspond to different stationary states attainable in the split dissipation regime. Thus, for different points on the phase diagram, the value of detuning is not the same. If several stationary states are identified, we choose the first state in the soliton existence range. The phase diagram is averaged over 3 realizations and the pump laser frequency is swept at the speed $\frac{1}{10}\kappa_0^2/2\pi$, corresponding to a change of frequency $\kappa_0$ every 10 photon lifetimes ($2\pi/\kappa_0$). 
\subsubsection{Dynamical states in split dissipation (\texorpdfstring{$\mathcal{PT}$}{PT}-broken) regime}
We differentiate four dynamical states in this regime: multi or single soliton in resonator 1 (S1, blue region), soliton crystal in resonator 1 (C1, cyan region), coexistence of periodic coherent structures in resonator 1 and soliton in resonator 2 (S2/C1, green region), and soliton in resonator 2 (S2, yellow region). The parameter regions enabling their generation are coloured on the phase diagram and their characteristic intracavity intensity profile are shown in the insets. White region refer to the absence of solitonic states. 

At weak inter-resonator coupling ($J<\frac{1}{2}\kappa_0$), the system qualitatively follows the single resonator dynamics and features the S1 state, where DKSs exist in resonator 1 while resonator 2 only features their low-power projection. Increasing the inter-resonator coupling, dynamical regions corresponding to states C1, S2/C1, and eventually S2 are accessed. 

The states are almost exactly partitioned in the resonator basis. For example, state S1 is confined in resonator 1 although a negligible amount ($\ll 1\%$) is found in resonator 2. That is, the field amplitude distribution between the resonators does not follow the supermode distribution which is given by linear analysis. We assume that the nonlinearity changes the field distribution of the supermode, making them localized in the resonators. These supermodes are referred to as high-loss (confined in resonator 1) and low-loss (confined in resonator 2).

\subsubsection{Parametric switching of the soliton-localisation}

In the range of $1.2 \kappa_0\lesssim J \lesssim 1.8\kappa_0$, the four stable states can also be accessed by changing the pump power. Fig.~\ref{fig:phase_space_case1}(b) shows the intracavity power evolution as a function of the laser detuning for pump power levels corresponding to four dynamical states in this range of $J$. Therefore, gradually increasing the pump power, states S2, S2/C1, C1, and S1 can be sequentially accessed.

In this process, the soliton-localization switches from resonator 2 to resonator 1. 
Linear analysis predicts that the resonant soliton will be confined in the low-loss supermode while the red-detuned CW background will be in the high-loss supermode [see Fig.~\ref{fig:criticalCouplingSimple}(c)] as is the case for state S2 [inset of Fig.~\ref{fig:phase_space_case1}(a)].
However, despite the field distribution predicted by the linear analysis, the presence of nonlinearity in the system introduces a mechanism allowing for the \emph{parametric switching} between the cavities.

Indeed, at low pump power, only the low-loss supermode has a quality factor sufficient for the soliton generation, resulting in state S2. At higher pump powers, both supermodes can sustain a coherent structure, leading to the coexistance of soliton and periodic coherent structure that has been observed in a limited intermediate range of parameters. Above a threshold, solitons are not generated in resonator 2. Moreover, in the C1 and S1 state, the parametric gain is able to compensate the difference of losses between the supermodes, and invert the $\mathcal{PT}$ symmetry: the parametric gain (via intra-band FWM) is larger in the supermode localized in resonator 1 than in the other supermode making the state of broken $\mathcal{PT}$-symmetry flipped in comparison to the linear regime for longitudinal modes with $\mu\neq 0$.

We note that no specific solitonic state was found at the EP. We suppose that the Kerr shift lifts the degeneracy between the two supermodes. However, an extensive investigation of the soliton generation in the close vicinity of the EP is beyond the scope of this study.

\subsection{\label{sec:311} Deterministic soliton crystal and efficient comb generation}

\begin{figure*}
	\centering
	\includegraphics[width=\linewidth]{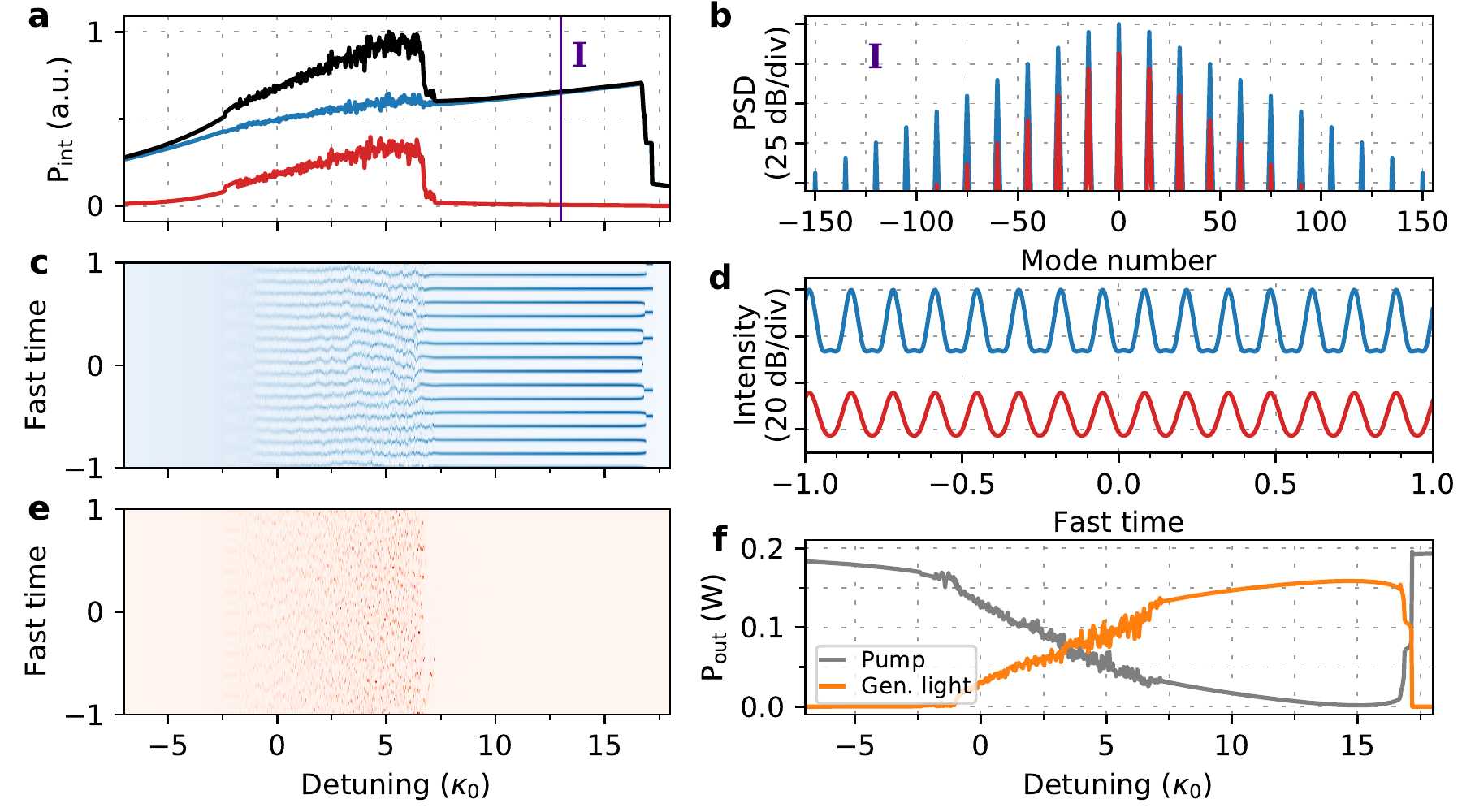}
	\caption{\textbf{Deterministic perfect soliton crystal generation in the $\mathcal{PT}$-symmetry broken phase. }(a) Intracavity power as a function of detuning. (b,d) Spectrum (b) and waveform (d) at detuning I. (c,e) Spatio-temporal diagrams for resonators 1 (c) and 2 (e). (f) Output of the through port, showing pump conversion above 75 \% into coherent comb. $P_\mathrm{in}=$0.2~W and $J/2\pi = $ 75~MHz.  Other parameters are identical to Fig.~\ref{fig:phase_space_case1}.}
	\label{fig:sim_C1}
\end{figure*}

In this section, we study the generation of state C1, which corresponds to the deterministic generation of a soliton crystal~\cite{karpov2019dynamics}. Fig.~\ref{fig:sim_C1} shows the numerical simulations of Eq.~\ref{eq:cLLEs} setting the pump power to 0.2~W and $J/2\pi = $ 75~MHz. Fig.~\ref{fig:sim_C1}(a) shows the intracavity power in both resonators as a function of laser detuning. The incident light couples into both high- and low-loss, supermodes of the system simultaneously, 
such that the low-loss supermode features a chaotic regime while the high-loss supermode remains in the cnoidal wave regime [Fig.~\ref{fig:sim_C1}(c,e)].
After passing a critical detuning ($\sim 7\kappa_0$), resonator 2 leaves the chaotic regime without any coherent structures generated while cnoidal waves of resonator 1 transition into a soliton crystal state. Fig.~\ref{fig:sim_C1}(b) shows that the crystal state at detuning I [Fig.~\ref{fig:sim_C1}(a)] is perfect~\cite{karpov2019dynamics} with more than 100 dB of extinction over almost the full existence range.

This state is known to exhibit a high conversion efficiency due to the high occupancy of the resonator 1 as shown in Fig.~\ref{fig:sim_C1}(b,d). Fig.~\ref{fig:sim_C1}(f) shows the output power in the pump mode ($\mu=0$) and comb modes ($\mu\neq0$). We observe that the perfect soliton crystal formation leads to a conversion efficiency higher than 75\%. Also, we note that the pump is almost completely absorbed by the system, such that an effective nonlinear critical coupling is achieved. 
According to~\cite{karpov2019dynamics}, the soliton crystal is generated deterministically when the pump power is below the threshold to avoid spatiotemporal chaos under the condition that modal crossings with higher-order modes trigger background modulation. Here, we observe deterministic soliton generation in the absence of modal crossings.

\subsection{\label{sec:313} Bright-dark solitons coexistence and their interaction with periodic coherent structures}
\begin{figure*}
	\centering
	\includegraphics[width=\linewidth]{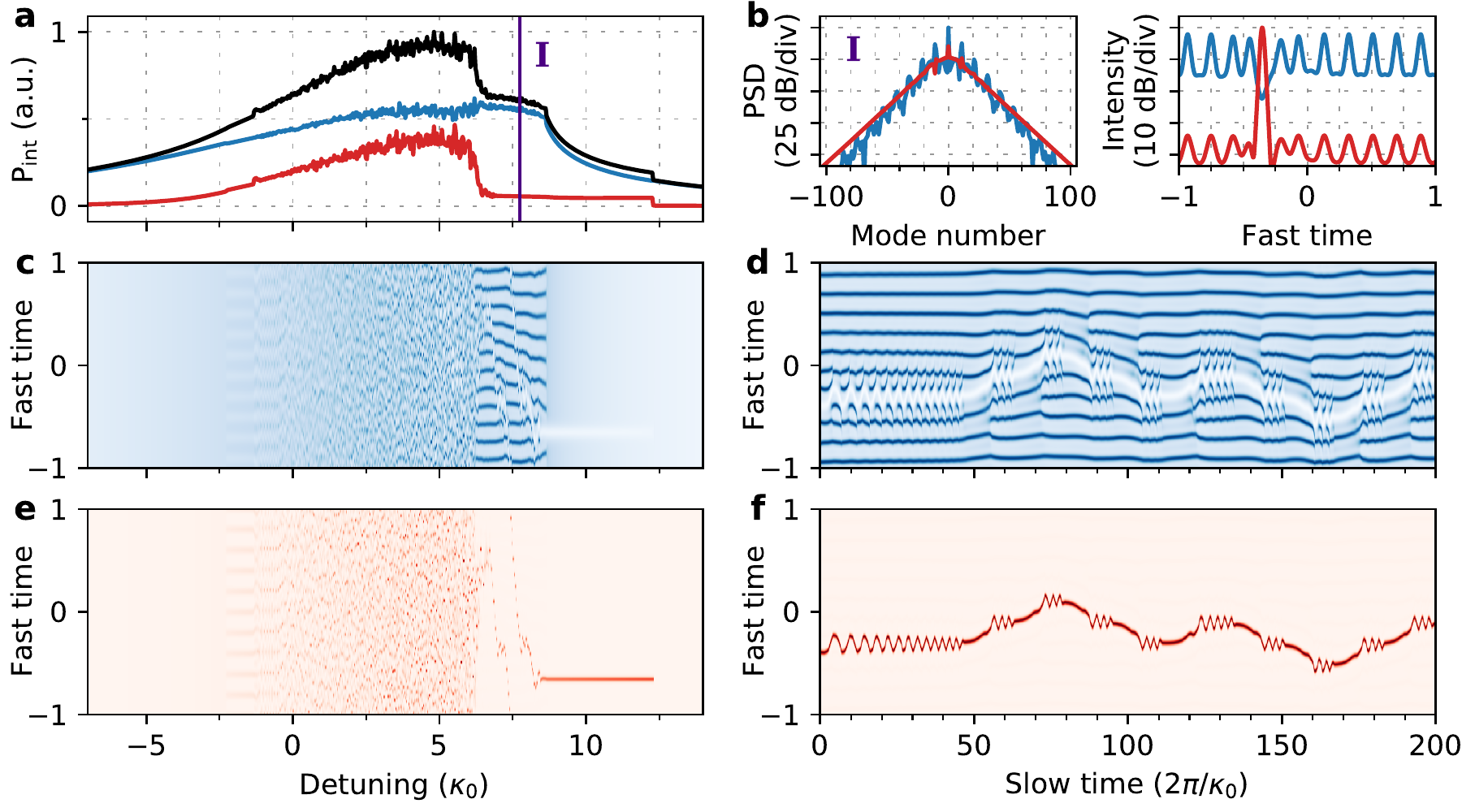}
	\caption{\textbf{Generation of bright-dark soliton pairs and interaction with periodic coherent structures in the $\mathcal{PT}$-symmetry broken phase.} (a) Intracavity power as a function of laser detuning. (b) Spectrum and waveform at detuning I. (c,e) Spatiotemporal diagram in resonator 1 (c) and 2 (e). (d,f) Simulation at fixed detuning seeded with the state from I, showing the evolution of the bright-dark soliton pair interacting with periodic coherent structures in resonator 1.  $P_\mathrm{in}=$0.1~W. Other parameters are identical to Fig.~\ref{fig:sim_C1}. }
	\label{fig:case1_S2C1}
\end{figure*}
We perform and analyse a simulation with $\Jc/2\pi = 75~\mathrm{MHz}$, $P_\mathrm{in} = 0.1~\mathrm{W}$, as shown in~Fig.~\ref{fig:case1_S2C1} in order to generate S2/C1 state depicted by green in the phase diagram [Fig.~\ref{fig:phase_space_case1}(a)]. The power trace [Fig.~\ref{fig:case1_S2C1}(a)] shows the presence of a step in each resonator. The spectrum and temporal intensity at detuning I are shown in Fig.~\ref{fig:case1_S2C1}(b). A soliton exists in resonator 2, while background modulation reminiscent of C1 state are present in both resonators.
The comb modes in both resonators are excited in this state, hinting at nonlinearly-induced $\mathcal{PT}$-transition that restores the $\mathcal{PT}$ symmetry in the comb modes~\cite{lumer2013nonlinearly,hassan2015nonlinear}. Spatiotemporal diagrams shown in Fig.~\ref{fig:case1_S2C1}(c,e) as a function of the laser detuning indicate that the S2/C1 state decays into a S2 state after the end of the soliton existence range in resonator 1. After transitioning to the S2 state, the field in resonator 1 acts as a source for resonator 2, resonantly supplying additional energy to the solitonic state. This results into a coexistence of a bright and dark solitons synchronously rotating in the resonators. This situation is similar to a dual fiber loop arrangement presented in~\cite{Xue2019Super} but in the limit of equal cavities.

While the existence of a periodic coherent structure in resonator 1, bright-dark soliton pair can be generated as well. A simulation at a fixed detuning starting from the initial conditions I [Fig.~\ref{fig:case1_S2C1}(b)] is shown in the spatiotemporal diagrams Fig.~\ref{fig:case1_S2C1}(d,f). We observe that the soliton pair is bounded by the effective nonlinear potential induced by the periodic structures in the neighbouring cavity. The spatiotemporal diagram depicts the possibility for the bright-dark soliton pair to tunnel from one potential unit cell to another interacting with their boundaries in an oscillatory manner. In addition to the fast oscillations, a random walk of the pair is observed at a slower timescale. 

\section{Discussion and conclusion}
\label{sec:disc_and_concl}

In this article we investigate nonlinear dynamics in a driven-dissipative photonic dimer exhibiting an exceptional point. We analyse the generation of dissipative Kerr solitons on both sides of the exceptional point which acts as a demarcation of the dimer critical coupling conditions. These two regimes are the split resonance regime (with the preserved symmetry) and split dissipation regime (with broken symmetry), as found in conventional $\mathcal{PT}$-symmetric systems with gain and loss. Each regime exhibits unique nonlinear dynamics not found in the single resonator. 


In the split resonance regime, which has been substantially discussed in~\cite{tikan2020emergent}, we observe that dimer solitons can be generated in either supermodes, however, only the antisymmetric one exhibits non-conventional soliton dynamics related to the emerging efficient four-wave mixing pathways. The dynamics is conveniently expressed in the supermode basis, for which we developed the concepts of inter-band four wave mixing. Supermode representation reveal that, despite the complexity of the dimer system, we are able to separate conventional single-resonator soliton dynamics from the dispersive waves emerging in another supermode. Breathing state of the photonic dimer in the supermode basis appears remarkably similar to its single-resonator counterpart except for a small perturbation. In this case, the intra-resonator power of both cavities oscillates in phase. 
Rapid and counter-phase power oscillations (soliton hopping) are observed above a threshold pump power, originating from the generation of synchronized solitons in both supermodes.
We highlight the fact that all the dynamics appearing in the split resonance regime can be well understood in the supermode representation.

The same does not apply to the regime of split dissipation. 
The absence of the resonance splitting implies the pumping of both supermodes simultaneously. Therefore, the most convenient representation in this case is the conventional resonator basis which exhibits the broken $\mathcal{PT}$-symmetry of the system. Satisfying the critical coupling condition, we impose different loss rates on the two resonators so resonator 1 becomes substantially overcoupled. 
We observe the generation of four different stable solitonic states localized in either or both resonators. In these states we observe: the synchronization of bright-dark soliton pairs (in resonators 2 and 1, respectively), the interaction of periodic coherent structures in resonator 1 with the bright-dark soliton pair, the deterministic generation of soliton crystal states with more than 75\% pump conversion efficiency into the comb lines, and bright solitons in resonator 1. 
Thereby, the pump power enables the parametric switching of the soliton localization between resonator 1 and 2. The switching is induced by flipping the broken $\mathcal{PT}$ symmetry so the lossy resonator 1 has more gain in the comb modes than resonator 2. We suspect nonlinearly induced $\mathcal{PT}$ transition to take place in the intermediate regime~\cite{lumer2013nonlinearly,hassan2015nonlinear}.
Moreover, we compute the nonlinear coupling coefficients between the supermodes. We normalize the divergence of two of them. One however seems to exhibit a singularity at the exceptional point, which could demonstrate enhanced sensitivity in its vicinity~\cite{Hodaei2017Enhanced}.

Concluding, we would like highlight the abundance of nonlinear dynamics occurring in the simplest element of soliton resonator lattices - photonic dimer. Despite the formal similarities with the single-mode dimer systems extensively studied in the context of non-Hermitian photonics, massively multimode nonlinear dimer exploiting another degree freedom reveals unprecedented solitonic states and dynamics, which cannot be covered in depth in one study. The fundamental aspects of this systems - two ideal coupled resonators - can be of interest far beyond the photonics community. 

\section{Acknowledgements}
This publication was supported by Contract18AC00032 (DRINQS) from the Defense Advanced Research Projects Agency (DARPA), Defense Sciences Office (DSO). This material is based upon work supported by the Air Force Office of Scientific Research under award number FA9550-19-1-0250. This work was further supported by the European Union’s Horizon 2020 Research and Innovation Program under the Marie Skłodowska-Curie grant agreements 846737 (CoSiLiS) and 812818 (MICROCOMB), and by the Swiss National Science Foundation under grant agreements 192293. Si$_3$N$_4$samples were fabricated and grown in the Center of MicroNanoTechnology (CMi) at EPFL.

\appendix

\section{\label{sec:appA} Definitions}
In this work, we define the Fourier Transform of a function $f$ as
\begin{equation}
    \FFT[f]_\mu=\frac{1}{2\pi}\int_0^{2\pi} d\theta f(\theta)e^{-i\mu\theta}.
\end{equation}
Thereby, the amplitude $A_\mu$ of the cavity mode with index $\mu$ is defined from the slowly-varying field amplitude co-rotating with angular frequency $D_1$ as 
\begin{equation}
    A_\mu = \FFT[A(\theta)]_\mu.
\end{equation}
This relation can be inverted to yield
\begin{equation}
A(\theta) = \sum_\mu A_\mu e^{i\mu\theta}.
\end{equation}
These definitions allows us to conveniently switch the nonlinear interaction from a temporal to a Fourier perspective~\cite{Hansson2014numerical}:
\begin{align}
     \FFT[A\abssq{A}]_\mu 
     = \sum_{\mu',\mu'',\mu'''}\!\!\!\!\!\!A_{\mu'}  A_{\mu''}  A^*_{\mu'''} \delta_{\mu+\mu''',\mu'+\mu''}
     \label{eq:nldecompose}.
\end{align}

\section{Critical coupling with inter-resonator detuning \label{sect:C}}
A non-vanishing value of inter-resonator detuning does not allow exceptional point (unless the coupling is complex). However, it is still important to find critical coupling conditions in the split resonance case in the general case including inter-resonator detuning. Indeed, two resonators will generally have different resonance frequencies such that $\delta\neq0$. Moreover, the use of integrated heaters \cite{tikan2020emergent} or piezo-electric components \cite{liu2020monolithic} can make the inter-resonator detuning a control parameter for the system.
Practically, a numerical method is easier to implement. Here, we would like to discuss qualitatively the dependence of the critical external coupling to the system parameters
 
In single resonators, critical coupling is achieved when the coupling rate matches the loss. This principle is maintained when more resonators are added. Indeed, Eq.~\ref{eq:critCoupl2} originates in the fact that each supermode is equally distributed in each resonator. Thus, the effective coupling to the through port and drop port are halved, whereas the intrinsic loss remains constant, because the circulating photons see the each port only half of the time. Equalizing the loss and the effective coupling between the waveguide and the supermode results in $\tfrac{1}{2}\kap1 = \kappa_0 + \tfrac{1}{2}\kap2$, as given by Eq.~\ref{eq:critCoupl2}. Similarly, the critical external coupling in the case of split dissipation is relatively high because coupling is achieved in the state with a low loss, which is only sparsely confined in resonator 1, such that the effective coupling is equal to the loss.
 
\begin{figure*}
    \centering
    \includegraphics[width=\linewidth]{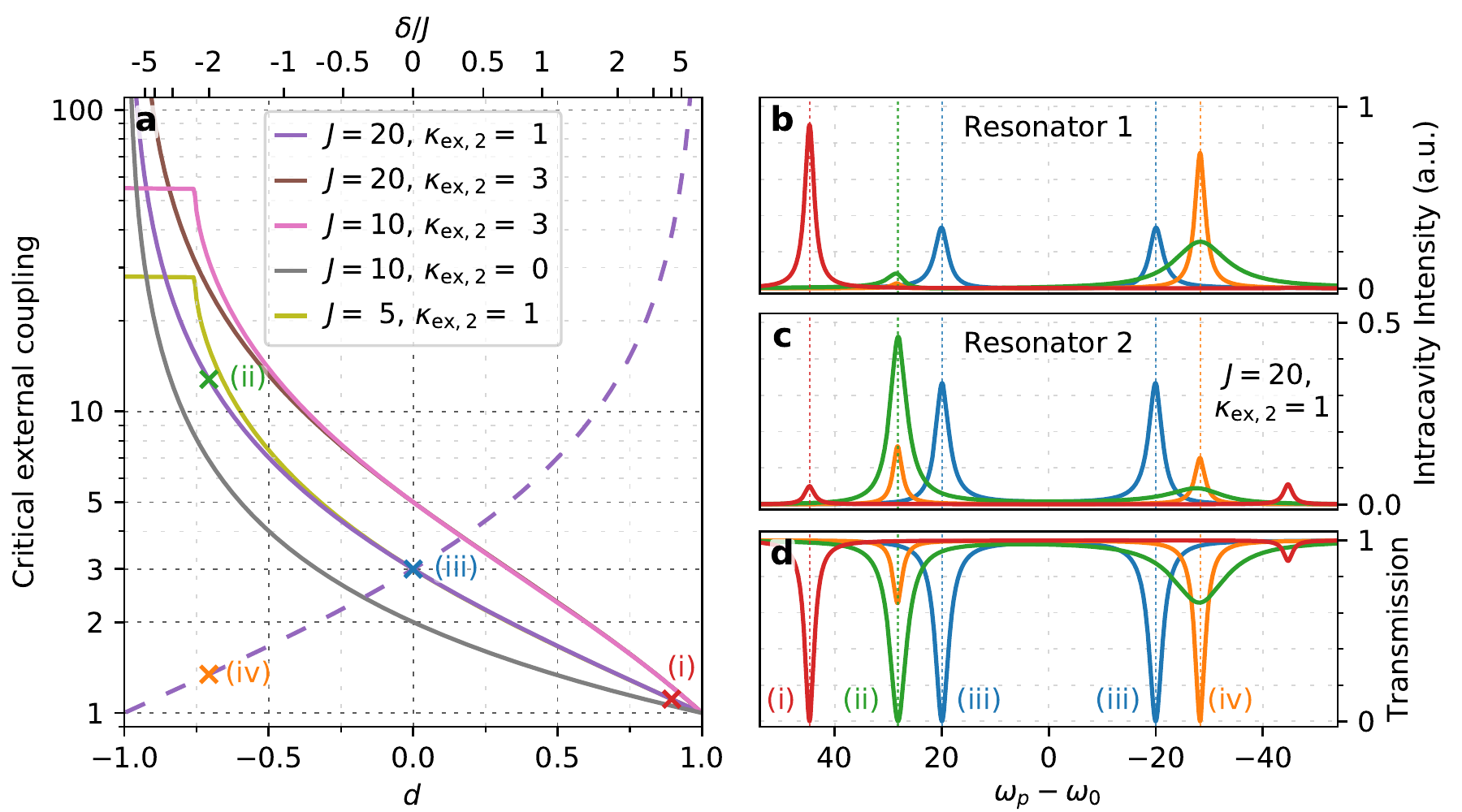}
    \caption{\textbf{Analysis of the critical coupling condition in the split resonance ($\mathcal{PT}$-symmetric) regime with non-vanishing inter-resonator detuning.} (a) Critical coupling condition to the AS resonance (solid) and to the S resonance (dashed) as a function of $d$. The corresponding ratio $\delta/\Jc$ is shown on the top. Plateaus indicate inaccessible critical coupling. (b-d) Representative intracavity fields and transmission in function of the laser detuning at four sets of parameters specified in (a). Parameters (iii) ($d=0$) allow critical coupling to both resonances. All units are normalized to $\kappa_0$. }
    \label{fig:criticalCouplingCase2}
\end{figure*}

When the inter-resonator detuning has a non-vanishing value, the supermodes are unequally distributed between the resonators. The effective coupling of each supermode to the through port depends on the hybridization, which strongly depends on $\delta$. For $\delta>0$, the AS supermode, which has a higher resonance frequency, is rather confined in resonator 1. Thus, the critical coupling condition is achieved for $\kap1 < 2\kappa_0 + \kap2$. Inversely, the S supermode requires higher $\kap1$ to achieve critical coupling. 
Fig.~\ref{fig:criticalCouplingCase2}(a) shows example curves of the critical coupling condition to the AS supermode for different parameters in split resonance regime as a function of the normalized inter-resonator detuning $d$. The dashed line shows the critical coupling condition to the S mode for one set of parameter. Four critical coupling conditions are marked and the intracavity fields in resonator 1, resonator 2, and the transmission spectrum at these sets of parameters are shown in Fig.~\ref{fig:criticalCouplingCase2}(b-d), respectively.

\section{\label{sec:appB}Projection into supermodes}
\subsection{Single longitudinal mode closed system}\label{subsect:closedSystem}
In order to analyze Eq. \ref{eq:cLLEs}, it is first simplified to a non-linear two level system isolated from the environment, that is
\begin{align}
\frac{\mathrm{d}}{\mathrm{d}t} A &= -i\left(\omega_0 + \frac{\delta}{2}\right) A + i \Jc B + i \gc A \abssq{A}\notag\\ 
\frac{\mathrm{d}}{\mathrm{d}t} B &= -i \left(\omega_0 -\frac{\delta}{2}\right) B + i \Jc A + i \gc B \abssq{B}.
\label{eq:closedEq}
\end{align}

For $\gc = 0$, the system can be diagonalized by defining the variables:
\begin{align}
\AS = \alpha A + \beta B \qquad & \AAS = \beta A - \alpha B\\
\alpha = \frac{\sqrt{1-d}}{\sqrt{2}} \qquad & \beta = \frac{\sqrt{1+d}}{\sqrt{2}},\label{eq:alphaBeta}
\end{align}
where $d = \delta/\Delta \omega$ and $\Delta \omega = \sqrt{4 \Jc^2+\delta^2}$, such that the Eq.~\ref{eq:closedEq} becomes 
\begin{align}
\frac{\mathrm{d}}{\mathrm{d}t}\AS &= -i\left(\omega_0 - \frac{\Delta \omega}{2}\right) \AS + i\gc ( \alpha A \abssq{A} + \beta B \abssq{B})\notag\\ 
\frac{\mathrm{d}}{\mathrm{d}t} \AAS &= -i\left(\omega_0+\frac{\Delta \omega}{2}\right) \AAS  + i\gc ( \beta A \abssq{A} - \alpha B \abssq{B}).
\label{eq:closedEq2}
\end{align}
We expand the non-linear part of Eq.~\ref{eq:closedEq2} using $A = \alpha \AS + \beta \AAS$ and $ B = \beta \AS -\alpha \AAS$. For the S mode, for example,
\begin{align}
 &\alpha A \abssq{A} + \beta B \abssq{B} = \notag\\
 & (\alpha^2 \abssq{\alpha}+ \beta^2 \abssq{\beta}) \AS\abssq{\AS}+ 2\alpha\beta(\abssq{\alpha}-\abssq{\beta} )\AAS\abssq{\AS} \notag \\
 &+ (\alpha^3 \beta^*-\beta^3 \alpha^*) \AS^2\AAS^*  +2(\alpha^2 \abssq{\beta} +\beta^2 \abssq{\alpha} )\AS \abssq{\AAS} \notag\\
 &+ (\beta^2\abssq{\alpha} + \alpha^2\abssq{\beta}) \AAS^2\AS^* +\alpha\beta(\abssq{\beta}-\abssq{\alpha})\AAS\abssq{\AAS}.
  \label{eq:nl1}
\end{align}
Therefore, Eq.~\ref{eq:closedEq2} can be rewritten as
\begin{align}
\frac{\mathrm{d}}{\mathrm{d}t} \AS &= -i(\omega_0-\frac{\Delta \omega}{2}) \AS + i\gc 
	\big[\tfrac{1}{2}(1+d^2) \AS\abssq{\AS} \notag \\
	&\quad-d \sqrt{1-d^2} \AAS\abssq{\AS} - \tfrac{1}{2}d\sqrt{1-d^2}  \AS^2\AAS^*\notag \\
 &\quad+(1-d^2)\AS \abssq{\AAS} + \tfrac{1}{2}(1-d^2) \AAS^2\AS^*\notag \\
 &\quad+ \tfrac{1}{2}d\sqrt{1-d^2} \AAS\abssq{\AAS} \big]  \notag \\ 
\frac{\mathrm{d}}{\mathrm{d}t} \AAS &= -i(\omega_0+\frac{\Delta \omega}{2}) \AAS  + i\gc  
	\big[ -\tfrac{1}{2}d\sqrt{1-d^2} \AS\abssq{\AS}\notag\\
	&\quad+ (1-d^2) \AAS\abssq{\AS} + \tfrac{1}{2}(1-d^2) \AS^2\AAS^*\notag\\
 &\quad+d\sqrt{1-d^2} \AS \abssq{\AAS} + \tfrac{1}{2}d\sqrt{1-d^2} \AAS^2\AS^*\notag\\
 &\quad+\tfrac{1}{2}(1+d^2)\AAS\abssq{\AAS}\big].
\label{eq:closedNLeq}
\end{align}


\subsection{Derivation with complex inter-resonator detuning} \label{subsect:openNonLinear}
The same analysis can be adapted when drive and dissipation are included. However, the inter-resonator detuning takes a complex component accounting for different loss rates between the resonators. 
We therefore start with
\begin{align}
\frac{\mathrm{d}}{\mathrm{d}t} A &= \left[-i(\omega_0 + \frac{\delta}{2})-\frac{\kappa_0+\kappa_{\mathrm{ex},1}}{2}\right] A + i \Jc B \notag\\
&\quad+ i \gc A \abssq{A} + \sqrt{\kappa_{\mathrm{ex},1}}s_\mathrm{in} \notag\\ 
\frac{\mathrm{d}}{\mathrm{d}t} B &= \left[-i (\omega_0 -\frac{\delta}{2})-\frac{\kappa_0+\kappa_{\mathrm{ex},2}}{2}\right] B + i \Jc A \notag\\
&\quad+ i \gc B \abssq{B},
\label{eq:openNLsys}
\end{align}
and define $\Delta \kappa_\mathrm{ex}$ to satisfy
\begin{align}
\kappa_{\mathrm{ex},1} = \kappa_\mathrm{ex} + \tfrac{1}{2}\Delta \kappa_\mathrm{ex} \notag\\
\kappa_{\mathrm{ex},2} = \kappa_\mathrm{ex} - \tfrac{1}{2}\Delta \kappa_\mathrm{ex},
\end{align}
such that we can define the complex inter-resonator detuning $\delta_c=\delta - i \frac{1}{2}\Delta \kappa_\mathrm{ex}$. We apply the following non-unitary transformation to Eq. \ref{eq:openNLsys}
\begin{align}
\AS &= \alpha A + \beta B &\AAS & = \beta A -\alpha B\\
\alpha & = \frac{\sqrt{1-d_c}}{\sqrt{2}} & \beta &= \frac{\sqrt{1+d_c}}{\sqrt{2}}
\end{align}	
with $d_c = \delta_c/\Delta\omega_c$ and $\Delta \omega_c = \sqrt{4J^2+\delta_c^2}$ to obtain
\begin{align}
\frac{\mathrm{d}}{\mathrm{d}t} \AS &= \left(-i\omega_0 - \frac{\kappa_0 + \kappa_\mathrm{ex}}{2}\right)\AS+ i \tfrac{1}{2}\Delta \omega_c \AS\notag\\
&\quad + i\gc \left( \alpha A \abssq{A} + \beta B \abssq{B}\right) +\alpha \sqrt{\kappa_{\mathrm{ex},1}}s_\mathrm{in}\notag\\ 
\frac{\mathrm{d}}{\mathrm{d}t} \AAS &= (-i\omega_0 - \frac{\kappa_0 + \kappa_\mathrm{ex}}{2}) \AAS -i \tfrac{1}{2}\Delta \omega_c \AAS \notag\\
&\quad+ i\gc ( \beta A \abssq{A} - \alpha B \abssq{B}) +\beta \sqrt{\kappa_{\mathrm{ex},1}}s_\mathrm{in}.
\end{align}
Here, the square roots of the complex numbers are taken as the root with positive real part as is the case when $\Delta\kappa_\mathrm{ex}=0$.
The complex inter-resonator detuning leads to a complex frequency splitting $\Delta\omega_c $, which can be approximated to second order in $\frac{\Delta\kappa_{ex}}{\Delta\omega}$ in the split resonance regime as 
\begin{align}
&\sqrt{4J^2 + (\delta - i \tfrac{1}{2}\Delta \kappa_\mathrm{ex})^2 } \notag\\
& \approx\quad\Delta\omega\left(1+\tfrac{1}{8}(d^2-1) \left(\frac{\Delta\kappa_{\mathrm{ex}}}{\Delta\omega}\right)^2\right) - i \tfrac{1}{2} \delta \frac{\Delta\kappa_{\mathrm{ex}}}{\Delta\omega}.
\end{align}
As in the closed case, $A = \alpha \AS + \beta \AAS,~B = \beta \AS - \alpha \AAS$ and the non-linear expansions such as Eq.~\ref{eq:nl1} are retrieved, such that 
\begin{align}
\frac{\mathrm{d}}{\mathrm{d}t} \AS &=\left[-i\left(\omega_0-\tfrac{1}{2}\Re(\Delta \omega_c)\right)\right.\notag\\
&\left.\qquad- \tfrac{1}{2}\left(\kappa_0 + \kappa_\mathrm{ex}+\Im(\Delta \omega_c)\right)\right]\AS +\alpha \sqrt{\kappa_{\mathrm{ex},1}}s_\mathrm{in} \notag\\
&\quad+ i\gc \left[t_1(d_c) \AS\abssq{\AS} + t_2(d_c) \AAS\abssq{\AS}\right. \notag\\
&\qquad   + t_4(d_c)  \AS^2\AAS^* +t_3(d_c)\AS \abssq{\AAS} \notag\\
&\qquad\left.+ \tfrac{1}{2}t_3(d_c) \AAS^2\AS^* - \tfrac{1}{2}t_2(d_c) \AAS\abssq{\AAS}\right]\notag\\ 
\frac{\mathrm{d}}{\mathrm{d}t} \AAS &= \left[-i\left(\omega_0+\tfrac{1}{2}\Re(\Delta \omega_c)\right)\right. \notag\\
&\qquad\left. - \tfrac{1}{2}\left(\kappa_0 + \kappa_\mathrm{ex}-\Im(\Delta \omega_c)\right)\right] \AAS +\beta \sqrt{\kappa_{\mathrm{ex},1}}s_\mathrm{in} \notag\\
&\quad+ i\gc \left[ \tfrac{1}{2}t_2(d_c) \AS\abssq{\AS}+ t_3(d_c)\AAS\abssq{\AS} \right. \notag \\
	 &\qquad  + \tfrac{1}{2}t_3(d_c) \AS^2\AAS^* -t_2(d_c) \AS \abssq{\AAS} \notag\\
	  & \left.\qquad- t_4(d_c) \AAS^2\AS^*+t_1(d_c)\AAS\abssq{\AAS} \right] 
\end{align}
with 
\begin{align}
t_1(d_c)&= \tfrac{1}{4}\left((1-d_c)\abs{1-d_c} + (1+d_c)\abs{1+d_c}\right) \notag\\
&\approx \tfrac{1}{2}(1+d^2) \notag\\
t_2(d_c) &= \tfrac{1}{2}\sqrt{1-d_c^2}\left(\abs{1-d_c}-\abs{1+d_c}\right)\notag\\
&\approx d\sqrt{1-d^2} \notag\\
t_3(d_c) &= \tfrac{1}{2}\left((1-d_c)\abs{1+d_c} + (1+d_c)\abs{1-d_c}\right)\notag\\
&\approx 1-d^2\notag\\
t_4(d_c) &= \tfrac{1}{4}\left((1-d_c)\sqrt{1-d_c+d_c^*-\abssq{d_c}} \right. \notag\\
&\quad \left.-(1+d_c)\sqrt{1+d_c-d_c^*-\abssq{d_c}}\right)\nonumber\\
&\approx -\frac{d}{2}\sqrt{1-d^2}.
\label{eq:def_nccs}
\end{align}
This derivation is valid in both the split dissipation and spit resonance regime. The approximations are considered for the latter regime, and are exact when $\Delta\kappa_{\mathrm{ex}} = 0$, where the result coincides with those found in Subsect.~\ref{subsect:closedSystem}. 

The same transformation can be applied for any pair of longitudinal mode, and the same coefficients are retrieved for the non-linear coupling terms. Due to the linearity of the Fourier transform, and in case of frequency independent inter-resonator detuning and coupling, it suffices to add the Fourier transform to the non-linear term to obtain Eq.~\ref{eq:nlcLLEs}.

\section{\label{sec:ApD}Solitonic state}

Solitonic state of the photonic dimer, as discussed in ~\cite{tikan2020emergent}, are accompanied by the periodic generation of DWs in the S supermode family. However, as follows from the Fig.~\ref{fig:sim_cc_2}(b), for certain values of the pump laser detuning the field in the S supermodes can be considered as a perturbation. Therefore, the system of coupled LLEs Eq.~\ref{eq:nlcLLEs} can be reduced to an effective single LLE in the supermode representation. 


Zones in the soliton existence range where the amplitude of the dispersive waves living in S supermode family is close to zero can be found in Fig.~\ref{fig:sim_cc_2}b, for example at the beginning of the green region.
Setting $\ASmu = 0,~\forall \mu$, Eq.~\ref{eq:nlcLLEs} becomes equivalent to the conventional equation for a single micro-resonator:
\begin{align}
\frac{\mathrm{d}}{\mathrm{d}t} \AASmu =& [-i(\omega_\mu- \mu D_1 -\omega_p+\tfrac{1}{2}\Re(\Delta \omega_c)) \notag\\
&\quad- \tfrac{1}{2}(\kappa_0 + \kappa_\mathrm{ex}-\Im(\Delta \omega_c))] \AASmu \notag\\
&\quad+\delta_{\mu,0}\beta \sqrt{\kappa_{\mathrm{ex},1}}s_\mathrm{in} + i t_1 \gc	\FFT[\AAS\abssq{\AAS}]_\mu,
\end{align}
 with rescaled effective parameters:
 \begin{align}
\gc^{(\mathrm{eff})} & = t_1 \gc\notag\\
\kappa_{\mathrm{ex}}^{(\mathrm{eff})} &=\beta^2 \kappa_{\mathrm{ex},1}\notag \\
\kappa_0^{(\mathrm{eff})} &=\kappa_0 - \tfrac{1}{2}\Delta\kappa_{\mathrm{ex}} - \Im(\Delta\omega_c) + \kappa_{\mathrm{ex},1}(1-\beta^2)\notag\\
\omega_0^{(\mathrm{eff})} &= \omega_0 + \tfrac{1}{2}\Re(\Delta \omega_c).
\label{eq:effective_parameters}
\end{align}
Hence, the validity of the vanishing S (AS) supermodes power implies that strongly coupled mulimode dimer can sustain conventional single resonator dissipative Kerr solitons in the AS (S) supermode family. The rescaling of the parameters accounts for the larger effective cavity volume if we consider the parameter $\gc$ for example.

\bibliography{apssamp}

\end{document}